\documentclass{aa}  

\usepackage{graphicx}
\usepackage{txfonts}
\usepackage{hyperref}
\hypersetup{
    colorlinks=true,
    linkcolor=blue,
    filecolor=magenta,      
    urlcolor=cyan,
    citecolor=blue
    }

\usepackage{graphicx}	
\usepackage{amsmath}	
\usepackage[version=4]{mhchem}
\usepackage{gensymb}
\usepackage{chemfig}
\usepackage{comment}

\begin{document}

   \title{Three-dimensional dynamical evolution of cloud particle microphysics in sub-stellar atmospheres}

   \subtitle{I. Description and exploring Y-dwarf atmospheric variability}

   \author{Elspeth K.H. Lee\inst{1} and Kazumasa Ohno\inst{2}}

   \institute{$^{1}$Center for Space and Habitability, University of Bern, Gesellschaftsstrasse 6, CH-3012 Bern, Switzerland \\
$^{2}$Division of Science, National Astronomical Observatory of Japan, 2-21-1 Osawa, Mitaka-shi, Tokyo, Japan}

    \authorrunning{Lee \& Ohno}

   \date{Received MM. DD, YY; accepted MM. DD, YY}

 
  \abstract
   {Understanding of cloud microphysics and the evolution of cloud structures in sub-stellar atmospheres remains a key challenge in the JWST era.
   The abundance of new JWST data necessitates models that are suitable for coupling with large-scale simulations, such as general circulation models (GCMs), in order to fully understand and assess the complex feedback effects of clouds on the atmosphere, and their influence on observed spectral and variability characteristics.}
   {We aim to develop a 2-moment, time-dependent bulk microphysical cloud model that is suitable for GCMs of sub-stellar atmospheres.
   }
   {We derive a set of moment equations for the particle mass distribution and develop a microphysical cloud model employing a 2-moment approach.
   We include homogeneous nucleation, condensation, and collisional microphysical processes that evolve the moments of a particle size distribution in time.
   We couple our new 2-moment scheme with the Exo-FMS GCM to simulate the evolution of KCl clouds for a WISE 0359-54 Y-dwarf parameter regime, and examine the effect of cloud opacity on the atmospheric characteristics.}
   {Our results show a global KCl cloud structure, with a patchy coverage at higher latitudes, as well as an equatorial belt region that shows increased particle sizes and variations in longitude. 
   Patchy regions are long lived, being present over many rotations of the brown dwarf.
   Our synthetic spectra conform well with JWST observations of WISE 0359-54, but more cloud opacity is required to dampen the spectral features at wavelengths below $\sim$7 $\mu$m.
   Our GCM shows periodic and sub-rotational variability on the order of 0.5-1\% in the Spitzer [3.6] and [4.5] micron bands, lower than that observed on other Y-dwarf objects.}
   {Our study demonstrates that the 2-moment bulk cloud microphysical scheme is a highly suitable method for investigating cloud characteristics and feedback in GCMs and other large scale simulations of sub-stellar atmospheres. 
   Split moment schemes and mixed material grains will be explored in a follow up study.}

   \keywords{Planets and satellites: atmospheres -- brown dwarfs -- Methods: numerical}

   \maketitle
%

\section{Introduction}

JWST has revolutionised both the quantity and detail of data available for characterising the atmospheres of exoplanets and brown dwarfs. 
In addition to offering clearer signatures of the gas phase compositions of sub-stellar atmospheres compared to previous data, JWST provides valuable constraints on the global cloud components in these atmospheres through measuring their wavelength dependent impacts on the spectrum.
This provides a valuable opportunity to model and investigate the complex three-dimensional (3D) processes driving cloud formation in these atmospheres.
It also enables a deeper understanding feedback mechanisms on thermal profiles, gas-phase chemistry, and atmospheric dynamics.
Unveiling how each microphysical process affects global cloud structures and goes on to impact observed properties of sub-stellar objects will be key questions to answer in the coming years.

Before JWST's launch, several studies looked at the possibility of detecting the infrared features of cloud particle absorption by different condensates for hot Jupiter exoplanets \citep{Wakeford2015}, as well as possibly inferring differences in the east-west limb-to-limb variance in cloud coverage \citep{Parmentier+16,Powell2019}.
Recently, evidence for these effects in JWST data have been discovered, with silicate features present in WASP-17b \citep{Grant2023}, WASP-107b \citep{Dyrek2024}, and HD 189733b \citep{Inglis+24} as well as cloud and chemistry limb asymmetry found in WASP-39b \citep{Espinoza2024} and WASP-107b \citep{Murphy+24}.

The evidence of limb asymmetry shows that cloud formation is inherently a 3D problem in gas giant atmospheres, with the microphysical properties of the clouds interacting with the dynamical flows of the atmosphere giving rise to this inhomogeneity.
\citet{Moran2024} suggest that the shape of the SiO$_{2}$ absorption features in hot Jupiters could infer the specific temperature-pressure dependent crystal structure of the cloud particles, further linking the local atmospheric thermochemical environment to the observed cloud properties.
Emission spectra are also aiding the characterisation of clouds, for example \citet{Bell2024} reported that the phase curve of WASP-43b showed consistent evidence of strong cloud coverage on the nightside of the planet.
\citet{Schlawin2024_WASP69b} also reported strong temperature and cloud inhomogeneity on the dayside of WASP-69b from its JWST emission spectrum.

Studies coupling time-dependent clouds in multi-dimensional simulations for hot Jupiter atmospheres have been explored in recent years, with several schemes available in the literature.
\citet{Lines+19} and \citet{Christie2021} couple the diagnostic eddysed model of \citet{Ackerman2001} to the Unified Model UK Met Office GCM to model the clouds on HD 209458b.
\citet{Komacek2022} present a single moment saturation adjustment scheme with relaxation timescale based on the model presented in \citet{Tan2019}.
This model was applied in an ultra hot Jupiter context in \citet{Komacek2022}, with similar saturation adjustment schemes now being used by several groups \citep[e.g.][]{Teinturier2024, Lee2024}.
\citet{Lee2016} and \citet{Lines2018} use a moment scheme based on the \citet{Helling2008} methodology coupled to a GCM to examine 3D cloud microphysics in the canonical hot Jupiters HD 189733b and HD 209458b.
\citet{Lee2023} present the `mini-cloud' 4-moment scheme based on the asymptotic giant branch outflow dust formation theory in \citet{Gail2013}, with additions for `dirty' mixed composition grains following \citet{Helling2008}.
\citet{Powell2024} used a pseudo-2D version of the CARMA bin microphysical cloud model \citep{Gao2018b,Gao2018}, showing a complex dynamic exchange of vapour and condensed species across the globe compared to static 1D models.
They show that differences in cloud composition and particle sizes are driven by dynamical processes as well as microphysical processes.
Despite this progress, modelling cloud microphysics self-consistently in exoplanet dynamical models remains a challenging prospect, but is key goal required for deeper understanding of the effect of clouds in hot Jupiter atmospheres.

In brown dwarf atmospheres, evidence of silicate features were seen in Spitzer IRAC data of L-dwarfs \citep[e.g.][]{Cushing2006, Suarez2022}, suggesting that active cloud formation was occurring in these objects.
In addition, photometric surveys using ground based instruments and Spitzer showed signs of variability in T and L dwarf atmospheres, in particular at the L-T transition \citep[e.g.][]{Radigan2014,Apai2017, Vos2019}, which suggested cloud formation was driving atmospheric variability.
HST observations, sometimes with combined with Spitzer, also provided evidence of cloud driven spectral variability in some objects \citep[e.g.][]{Yang2015, Biller2018, Bowler2020, Zhou2020}, with large $\gtrsim$ 10\% variations in the amplitude of the near-infrared \ce{H2O} features found in a variety of objects.

With this evidence of the effects of cloud coverage in brown dwarfs, the launch of JWST offers the ability to fully characterise the complexity of these clouds to understand their compositions, effect on the thermal and mixing properties of the atmosphere, and how they interact with chemical species.
\citet{Miles2023} report a wide wavelength range spectrum of VHS 1256 b, finding disequilibrium chemistry and a silicate feature.
\citet{Biller2024} recently published spectral variability light curves across the NIRSpec Prism and MIRI Low Resolution Spectrograph (LRS) wavelength ranges for WISE 1049A and WISE 1049B, finding evidence for chemical species and cloud coverage variability in their atmospheres, in addition to a variable silicate absorption feature for WISE 1049A.
This suggests a strongly inhomogenous atmosphere with dynamically driven small and large scale dynamical features forming and dissipating in these atmospheres at shorter timescales than the rotational period.

For Y-dwarf atmospheres, variability studies has thus far been performed using Spitzer and ground based instruments.
\citet{Cushing2016} observed WISE J140518.39+553421.3 ($T_{\rm eff}$ $\approx$ 350-400 K, $P_{\rm rot}$ $\approx$ 8.5 hrs) in the Spitzer [3.6] and [4.5] bands, finding rotational variability of $\approx$3.5\%.
\citet{Leggett2016} observed WISEP J173835.52+273258.9 ($T_{\rm eff}$ $\approx$ 425 K, $P_{\rm rot}$ $\approx$ 6.0 hrs) using ground based and Spitzer photometry, finding variability on the order of $\approx$1.5\% in the [4.5] Spitzer band and $\sim$5-15\% in the Y band.
Both the above studies note the need for 3D complex modelling to understand the origins of this variability.
{With JWST, several Y-dwarf objects now have spectral data across a wide wavelength regime \citep[e.g.][]{Beiler2023,Lew2024,Luhman2024,Beiler2024}.
This has lead to precise and in-depth characterisation of the chemical composition of atmospheres in this regime.
More JWST spectral characteristion and variability surveys are underway (\#2124: PI Faherty, \#2327: PI Skemer), which will hopefully uncover the variable, time-dependent chemical and cloud characteristics of a wide variety of brown dwarfs across the Y-dwarf parameter range.

Previous theoretical and modelling studies of brown dwarfs have revealed the time-dependent connection between atmospheric temperatures, dynamics, cloud coverage and chemistry.
\citet{Freytag2010} examined the mixing and radiative effects of clouds in brown dwarf atmospheres using 2D hydrodynamic modelling.
\citet{Zhang2014} used a pulsing thermal perturbation scheme to imitate the temperature forcing induced by convection near the radiative-convective boundary coupled to a global shallow-water model.
They found the strength of the atmospheric temperature variability greatly depended on the dynamical regime of the atmosphere.
\citet{Robinson2014} used a 1D model with deep temperature perturbations to investigate the connection between deep convective motions and the observed spectral features.
\citet{Morley2014b} modeled variability of T- and Y-dwarfs with patchy sulfide and salt cloud layers.
\citet{Tan2019} find that cloud opacity feedback can trigger convective motions, giving rise to cloud condensation and evaporation cycles.
\citet{Tremblin2020} apply modified temperature-pressure (T-p) profiles to mimic thermal variability and also investigate the effects of non-equilibrium chemistry on the variability and strength of spectral features.
\citet{Luna2021} use the \citet{Ackerman2001} equilibrium cloud model to produce vertical cloud profiles that attempt to fit the Spitzer IRS silicate feature in brown dwarfs \citep{Cushing2006,Looper2008} and Spitzer photometric band variability \citep{Metchev2015}.
\citet{Hammond2023} found four main dynamical regimes for brown dwarfs, depending on the non-dimensional radiative timescale and thermal Rossby number.
\citet{Lacy2023} show significant effects when considering the impact of \ce{H2O} clouds on Y-dwarf thermal structures.
\citet{Lee2023b} and \citet{Lee2024} examine the impact of large scale dynamical features in brown dwarfs and directly imaged exoplanets on the 3D chemical, cloud and variability characteristics of the atmosphere. 

Owing to the continued release of JWST data starting to characterise the cloud structures of sub-stellar atmospheres in detail, microphysical models are warranted to deeply understand the processes that give rise to the specific features and effects seen in the spectra of sub-stellar objects.
Bulk, also known as moment, schemes remain a popular option in Earth Science for modelling cloud microphysics due to their relative simplicity and computational efficiency \citep[for review, see e.g.,][]{Morrison+20_Earth_cloud_review}.
Only a few moments (in this paper, two) are required to be evolved in the dynamical system, with each moment representing an average value of the particle size distribution and the microphysical processes altering these mean values in time.

For sub-stellar atmospheres, bulk schemes have been developed by a few groups \citep[e.g.][]{Helling2008, Ohno2017,Ohno2018, Ormel&Min2019,Woitke2020, Lee2023,Huang2024} but only a few studies have used bulk schemes directly coupled time-dependently to 3D atmospheric simulations \citep[e.g.][]{Lee2016, Lines2018, Lee2023}.
Bulk schemes are contrasted with bin or spectral models \citep[e.g. CARMA,][]{Gao2018b}, which evolve the discretised particle size distribution in time through calculating the fluxes in and out each bin size.
Typically 40-70 or more mass or size bins are required to adequately resolve the size distribution in these models \citep[e.g.][]{Adams2019}, making them much more computationally demanding than bulk methods and more challenging to include in large scale 3D atmospheric modelling with present resources.

In this study, we expand the mini-cloud code package \citep{Lee2023} to include a new 2-moment cloud microphysical scheme able to be efficiently coupled to time-dependent sub-stellar atmosphere simulations in a simple manner.
The present 2-moment scheme also newly implements the collision growth of cloud particles, which is a key process to produce rain droplets in Earth clouds \citep{Morrison+20_Earth_cloud_review}. 
In Section \ref{sec:mom}, we present the mass moment scheme and detail the individual microphysical source and sink terms in the set of ordinary differential equations (ODEs).
In Section \ref{sec:GCM}, we couple our new 2-moment scheme to the 3D Exo-FMS GCM to simulate the Y-dwarf WISE 0359-54 as a test application.
In Section \ref{sec:postpro}, we post-process our simulation and generate emission spectra and variability light curves.
In Section \ref{sec:disc}, we discuss our results in context to other modelling efforts and observational data.
Section \ref{sec:conc} contains the summary and conclusions of our study.
Each of the new moment schemes are available online\footnote{\url{https://github.com/ELeeAstro/mini_cloud}}.

\section{Bulk microphysical scheme}
\label{sec:mom}

For this study, we consider the mass moments of the particle mass distribution.
The mass moments, M$^{(k)}$ [g$^{k}$ cm$^{-3}$], of the particle mass distribution are given by
\begin{equation}
    M^{(k)} = \int_{0}^{\infty}m^{k}f(m)dm,
\end{equation}
where k is the integer moment power, $m$ [g] the mass of a particle in the distribution and $f(m)$ [cm$^{-3}$ g$^{-1}$] the particle mass distribution.

Examining the moments up to the second (k = 2), leads to a set of moments each with physical meanings that represent bulk values of the particle size distribution:
\begin{equation}
    M^{(0)} = N_{\rm c},
\end{equation}
in units of cm$^{-3}$ which represents the total number density of the size distribution, 
\begin{equation}
  M^{(1)} = \rho_{\rm c},
\end{equation}
in units of g cm$^{-3}$ which represents the total mass density of the size distribution, and
\begin{equation}
  M^{(2)} = Z_{\rm c},
\end{equation}
in units of g$^{2}$ cm$^{-3}$ which is related to the Rayleigh regime radar reflectivity of the size distribution\footnote{In other words, $Z_{\rm c}$ is related to scattering efficiency of clouds. The scattering cross section of a cloud particle obeys $\sigma_{\rm Ray}\propto a^{6}\propto m^2$ in the Rayleigh regime \citep[e.g.,][]{Bohren1983}. This yields a scattering efficiency $=\int \sigma_{\rm Ray}f(m)dm$ being proportional to $Z_{\rm c}$.}.
$Z_{\rm c}$ is typically divided by the square of the atmospheric density $\rho_{\rm a}$ [g cm$^{-3}$], giving it the more regular units of cm$^{6}$ cm$^{-3}$.
Here the subscript c stands for `cloud' particles, but different sub-scripts can be used to distinguish various aerosol types such as haze \citep[e.g.][]{Ohno2024}.
Note that the mass moments have equivalents when taking moments of the radius distribution \citep[e.g.][]{Gail2013}, $N_{\rm c}$ is the zeroth moment, $\rho_{\rm c}$ the third moment and $Z_{\rm c}$ the sixth moment in radius space respectively.

Ratios and relationships between each of the moments provides information on the characteristics of the particle size distribution, for example the mean mass of a particle, $m_{\rm c}$ [g], is given by the ratio of the first to zeroth moment
\begin{equation}
\label{eq:m_c}
    m_{\rm c} = \frac{\rho_{\rm c}}{N_{\rm c}},
\end{equation}
and the average particle volume, $V_{\rm c}$ [cm$^{3}$],
\begin{equation}
\label{eq:V_c}
    V_{\rm c} = \frac{m_{\rm c}}{\rho_{\rm d}},
\end{equation}
where $\rho_{\rm d}$ [g cm$^{-3}$] is the bulk mass density of the particle, which is not necessary the material mass density if the particle porosity is taken into account \citep[e.g.,][]{Ohno2020}.
The mass or volume weighted mean particle radius, $r_{\rm c}$ [cm], can be derived as
\begin{equation}
\label{eq:r_c}
    r_{\rm c} = \sqrt[3]{\frac{3m_{\rm c}}{4\pi\rho_{\rm d}}}.
\end{equation}
The standard deviation, $\sigma_{\rm c}$ [g], of the mass distribution can be characterised through \citep[e.g.][]{Gail2013}
\begin{equation}
  \sigma_{\rm c} = \sqrt{\frac{Z_{\rm c}}{N_{\rm c}} - \left(\frac{\rho_{\rm c}}{N_{\rm c}}\right)^{2}}.
\end{equation}

Following the general method for deriving the time derivatives of the moments \citep[e.g.][]{Gail2013}, Appendix \ref{app:b}, the set of differential equations for the 2-moment scheme used in this study are given by
\begin{equation}
    \frac{dN_{\rm c}}{dt} = (J_{\rm hom} + J_{\rm evap}) + \left(\frac{dN_{\rm c}}{dt}\right)_{\rm coll},
\end{equation}
\begin{equation}
    \frac{d\rho_{\rm c}}{dt} = m_{\rm seed}(J_{\rm hom} + J_{\rm evap}) + \left(\frac{dm}{dt}\right)_{\rm cond}N_{\rm c},
\end{equation}
where $J_{\rm hom}~{\rm[cm^{-3}~s^{-1}]}$ is the homogeneous nucleation rate of seed particles, $J_{\rm evap}~{\rm[cm^{-3}~s^{-1}]}$ is the prescribed disappearance rate of seed particles through evaporation, and $m_{\rm seed}$ [g] is the mass of a seed particle.
The change in condensable vapour mass density, $\rho_{\rm v}$ [g cm$^{-3}$], is given by
\begin{equation}
    \frac{d\rho_{\rm v}}{dt} = -m_{\rm seed}(J_{\rm hom} + J_{\rm evap})- \left(\frac{dm}{dt}\right)_{\rm cond}N_{\rm c} + \left(\frac{d\rho_{\rm v}}{dt}\right)_{\rm deep}.
\end{equation}
In the current 2-moment scheme, we have assumed a monodisperse size distribution that can be uniquely identified by $N_{\rm c}$ and $\rho_{\rm c}$ (Appendix \ref{app:b}).
Thus, we ignore the second moment, $Z_{\rm c}$, and only evolve $N_{\rm c}$, $\rho_{\rm c}$ and $\rho_{\rm v}$ with time.
The source and sink terms are detailed in the following sections.

\subsection{Collisional growth processes}

Collisional growth rates are primarily governed by the Smoluchowski equation \citep{Smoluchowski1916} in its continuous form \citep{Muller1928}
\begin{equation}
\begin{split}
\label{eq:Smoluchowski}
    \frac{\partial n(m)}{\partial t} = & -n(m)\int_{0}^{\infty}K(m,m')n(m')dm' \\ &+ \frac{1}{2}\int_{0}^{m}K(m - m',m')n(m - m')n(m')dm',
\end{split}
\end{equation}
where $m$ [g] and $m'$ [g] are discrete particle masses of the distribution, $n(m)$ [cm$^{-3}$] the number density of particles of mass $m$ and $K$ [cm$^{3}$ s$^{-1}$] the collisional probability rate between two particle masses, typically called the collision kernel.

In \citet{Drake1972}, the mass moment generative function of the Smoluchowski equation is derived as
\begin{equation}\label{eq:coll_mom}
\begin{split}
    \frac{dM^{(k)}}{dt} = &\frac{1}{2}\int_{0}^{\infty}\int_{0}^{\infty}K(m,m')f(m)f(m') \\ & \times \left[(m + m')^{k} - m^{k} - m'^{k}\right]dmdm',
\end{split}
\end{equation}
where $m$ and $m'$ represent different masses in the particle mass distribution.
We can immediately get expressions for the zeroth and first moments, $N_{\rm c}$ and $\rho_{\rm c}$, for $k$ = 0 and 1 respectively
\begin{equation}
\label{eq:N_coll}
    \frac{dN_{\rm c}}{dt} = -\frac{1}{2}\int_{0}^{\infty}\int_{0}^{\infty}K(m,m')f(m)f(m')dmdm',
\end{equation}
\begin{equation}
    \frac{d\rho_{\rm c}}{dt} = 0.
\end{equation}
These rate expressions follow the expected behaviour for collisional growth interactions. 
For the zeroth moment, collisional growth is a sink term, reducing the number density with time and total mass is conserved during collisions, recovering a zero rate for the first moment.

The total rate of change of the zeroth moment given by collisional growth is the combination of the coagulation and coalescence processes
\begin{equation}
    \left(\frac{dN_{\rm c}}{dt}\right)_{\rm coll} = \left(\frac{dN_{\rm c}}{dt}\right)_{\rm coag} + \left(\frac{dN_{\rm c}}{dt}\right)_{\rm coal}.
\end{equation}
Throughout this study we assume `hit-and-stick' collisions, ignoring other effects such as bouncing and fragmentation.
We do not include turbulence driven collisional processes in this study, such as those considered in \citet{Samra2022}, since the energy dissipation rate of turbulence remains highly uncertain for brown dwarfs and exoplanet atmospheres.

\subsubsection{Brownian coagulation}

In collisional growth via Brownian motion (coagulation), particles interact via random walk, the rate of which depend on the local atmospheric properties and particle characteristics.
This is primarily dependent on the Knudsen number, Kn, regime 
\begin{equation}
    {\rm Kn} = \frac{\lambda_{\rm a}}{r_{\rm c}},
\end{equation}
where $\lambda_{\rm a}$ [cm] is the mean free path of the atmosphere given by
\begin{equation}
    \lambda_{\rm a} = \frac{2\eta_{\rm a}}{\rho_{\rm a}}\sqrt{\frac{\pi\bar{\mu}_{\rm a}}{8RT}},
\end{equation}
where $\bar{\mu_{\rm a}}$ [g mol$^{-1}$] is the local atmospheric mean molecular weight, $\rho_{\rm a}$ [g cm$^{-3}$] the mass density of the atmosphere and the dynamical viscosity, $\eta_{\rm a}$ [g cm$^{-1}$ s$^{-1}$], of the atmosphere given by the \citet{Rosner2012} fitting function
\begin{equation}
\label{eq:rosner}
\eta_{\rm a} = \frac{5}{16}\frac{\sqrt{\pi m k_{b}T}}{\pi d^{2}}\frac{(k_{b}T/\epsilon_{\rm LJ})^{0.16}}{1.22},
\end{equation}
where the parameters for the molecular diameter, $d$ [cm], mass, $m$ [g], and Lennard-Jones potential, $\epsilon_{\rm LJ}$, for \ce{H2}, He and H, the main components of hydrogen-rich atmospheres are listed in \citet{Lee2023}, though in mini-cloud we include values taken from \citet{Rosner2012} to account for the gas species used in the mini-chem chemical kinetics scheme \citep{Tsai2022}.

The dynamical viscosity of mixtures of gas species can be calculated from weighting the dynamical viscosity of each species with their local volume mixing ratios following the approximate classical mixing law from \citet{Wilke1950} or the approximate square root mixing law from \citet{Herning1936}.
In this study, we apply the more complex mixing law from \citet{Davidson1993}, which takes into account the momentum exchange between gas species \citep[e.g.][]{Gao2023}.

To derive the set of coagulation rate equations, we follow the arguments in \citet{Rossow1978} for similar sized particle-particle collisions at the mean particle size, $r_{\rm c}$.
\citet{Rossow1978} considers the coagulation collision kernel expression from \citet{Fuchs1964} but here we consider the kernel following \citet{Chandrasekhar1943} valid in the Kn $\ll$ 1 regime
\begin{equation}
    K = 4\pi\left[D(r)  + D(r')\right](r + r'),
\end{equation}
where $D(r)$ is the particle diffusion factor given by \citep{Chandrasekhar1943}
\begin{equation}
    D(r) = \frac{k_{\rm b}T\beta}{6\pi\eta_{\rm a}r},
\end{equation}
where $\beta$ is the Cunningham slip correction factor \citep{Davies1945}
\begin{equation}
    \beta = 1 + {\rm Kn} \left[1.275 + 0.4 \exp(-1.1/{\rm Kn})\right].
\end{equation}
Assuming similar sized collisions ($r$ $\approx$ $r'$, $m$ $\approx$ $m'$) at the mean particle size $r_{\rm c}$, applying this kernel expression into Eq. \ref{eq:N_coll} gives
\begin{equation}
\label{eq:N_coag1}
    \left(\frac{dN_{\rm c}}{dt}\right)_{\rm coag} \approx - \frac{4k_{\rm b}T\beta}{3\eta_{\rm a}}N_{\rm c}^{2}.
\end{equation}

For the free molecular flow or kinetic regime (Kn $\gg$ 1), a Maxwell-Boltzmann thermal velocity distribution can be assumed for the particles.
\citet{Rossow1978} uses the \citet{Brock1965} collision kernel but to facilitate clearer calculations we follow the \citet{Vincenti1965} kernel given by 
\begin{equation}
    K = (r + r')^{2}\sqrt{\frac{8\pi k_{\rm b} T}{\widetilde{m}}},
\end{equation}
where $\widetilde{m}$ = $m$$m'$/($m$ + $m'$) [g] is the reduced mass.
Again, assuming similar sized collisions ($r$ $\approx$ $r'$, $m$ $\approx$ $m'$) at the mean particle size $r_{\rm c}$ and putting this kernel into Eq. \ref{eq:N_coll} yields
\begin{equation}
\label{eq:N_coag2}
    \left(\frac{dN_{\rm c}}{dt}\right)_{\rm coag} \approx - 8\sqrt{\frac{\pi k_{\rm b} T}{m_{\rm c}}}r_{\rm c}^{2}N_{\rm c}^{2}.
\end{equation}
Overall, Eqs. \ref{eq:N_coag1} and \ref{eq:N_coag2} are same expressions found in \citet{Ohno2018} and \citet{Ohno2020}, which we here derived directly from the \citet{Drake1972} formulation.

An approach including an interpolation function between the continuum and kinetic regimes is also commonly used in the literature \citep[e.g.][]{Fuchs1964, Pruppacher1978, Lavvas2010, Gao2018b}.
In this case, the coagulation kernel is given by \citep{Gao2018b}
\begin{equation}
  K = \frac{4\pi\left[D(r)  + D(r')\right](r + r')}{\frac{r + r'}{r + r' + \sqrt{\delta_{r}^2 + \delta_{r'}^2}} + \frac{4\left[D(r)  + D(r')\right]}{(r + r')\sqrt{V_{r}^2 + V_{r'}^2}}},
\end{equation}
where $V_{r}$ is
\begin{equation}
    V_{r} = \sqrt{\frac{8k_{\rm b}T}{\pi m_{\rm r}}},
\end{equation}
and $\delta_{r}$ is
\begin{equation}
    \delta_{r} = \frac{(2r + \lambda_{r})^{3} - (4r^{2} + \lambda_{r}^{2})^{3/2}}{6r\lambda_{r}} - 2r,
\end{equation}
where $\lambda_{r}$
\begin{equation}
    \lambda_{r} = \frac{8D(r)}{\pi V_{\rm r}}.
\end{equation}
$\delta_{r}$ is the function that interpolates between the continuum (Eq. \ref{eq:N_coag1}) and kinetic (Eq. \ref{eq:N_coag2}) regimes.

Again, assuming similar sized collisions, $r$ $\approx$ $r'$ at the mean particle size $r_{\rm c}$, this kernel reduces to
\begin{equation}
    K = \frac{16\pi D(r_{\rm c})r_{\rm c}}{\frac{2r_{\rm c}}{2r_{\rm c} + \sqrt{2}\delta_{r_{\rm c}}} + \frac{4D(r_{\rm c})}{r_{\rm c}\sqrt{2}V_{r_{\rm c}}}}.
\end{equation}
Defining the correction factor as 
\begin{equation}
    \varphi = \frac{2r_{\rm c}}{2r_{\rm c} + \sqrt{2}\delta_{r_{\rm c}}} + \frac{4D(r_{\rm c})}{r_{\rm c}\sqrt{2}V_{r_{\rm c}}},
\end{equation}
and putting this kernel into Eq. \ref{eq:N_coll} gives
\begin{equation}
\label{eq:N_coag3}
    \left(\frac{dN_{\rm c}}{dt}\right)_{\rm coag} \approx - \frac{4k_{\rm b}T\beta}{3\eta_{\rm a}\varphi}N_{\rm c}^{2}.
\end{equation}
Due to the versatility of the interpolation function approach, we primarily make use of Eq. \ref{eq:N_coag3} in mini-cloud.

\subsubsection{Gravitational coalescence}

When there is a difference in the settling velocity between two different particle sizes, the larger particle settles faster, impacting and coalescing with the slower moving particles below.
The kernel for gravitational coalescence is given by \citep[e.g.][]{Jacobson2005}
\begin{equation}
    K = \pi(r + r')^{2}|v_{\rm f} (r) - v_{\rm f} (r')|E,
\end{equation}
where $r'$ $>$ $r$, $v_{\rm f} (r)$ [cm s$^{-1}$] the settling velocity and $E$ the collisional efficiency factor.
Defining the differential settling velocity as $\Delta v_{\rm f} $ = $|v_{\rm f} (r) - v_{\rm f} (r')|$ and assuming similar particle size collisions ($r$ $\approx$ $r'$) at the mean particle size $r_{\rm c}$, the differential equation for the zeroth moment is, following Eq. \ref{eq:N_coll},
\begin{equation}
  \left(\frac{dN_{\rm c}}{dt}\right)_{\rm coal} \approx -2\pi r_{\rm c}^{2}\Delta v_{\rm f}  E N_{\rm c}^{2}.
\end{equation}

Following \citet{Ohno2017} to evaluate $\Delta v_{\rm f} $ we introduce a parameter, $\epsilon$, that estimates the relative velocity of the particles from the mean particle size settling velocity, $v_{\rm f}(r_{\rm c})$, giving $\Delta v_{\rm f}$ $\approx$ $\epsilon$$v_{\rm f}(r_{\rm c})$. 
This is taken as $\epsilon$ $\approx$ 0.5 following the results of \citet{Sato2016}, who found this value to best reproduce the results of a collisional bin resolving model for grain growth in protoplanetary disks.
The settling velocity is given by the expression in \citet{Ohno2018}
\begin{equation}
\label{eq:vf}
    v_{\rm f}(r_{\rm c}) = \frac{2\beta gr_{\rm c}^{2}\rho_{\rm d}}{9\eta_{\rm a}}\left[1 + \left(\frac{0.45gr_{\rm c}^{3}\rho_{\rm a}\rho_{\rm d}}{54\eta_{\rm a}^{2}}\right)^{2/5}\right]^{-5/4},
\end{equation}
where $g$ [cm s$^{-2}$] is the gravity.
$E$ is a collisional efficiency factor, dependent on the Stokes number, Stk, 
\begin{equation}
    {\rm Stk} = \frac{\Delta v_{\rm f} v_{\rm f}(r_{\rm c})}{gr_{\rm c}} = \frac{\epsilon v_{\rm f}^{2}(r_{\rm c})}{gr_{\rm c}}.
\end{equation}
$E$ is then given by \citep{Guillot2014}
\begin{equation}
     E =
    \begin{cases}
      {\rm max}\left[0, 1 - 0.42{\rm Stk}^{-0.75}\right]. & {\rm Kn} < 1 \\
      1. &  {\rm Kn} \geq 1 
    \end{cases}
\end{equation}

\subsection{Condensation and evaporation}

Condensation occurs when a condensable vapour species is supersaturated with respect to its saturation vapour pressure, while evaporation occurs when the vapour is under-saturated.
In the small Knudsen number regime Kn $\ll$ 1, the growth of the particle is limited by the diffusion of condensable gas to the surface of the grain.
In this regime, we follow the derivation found in \citet{Woitke2003} but defined for mass instead of volume
\begin{equation}
\label{eq:cond_d}
    \left(\frac{dm}{dt}\right)_{\rm cond} = 4\pi r_{\rm c} D m_{0} n_{\rm v} \left(1 - \frac{1}{S}\right),
\end{equation}
where $S(T)$ = $p_{\rm par}/p_{\rm vap}(T)$ is the supersaturation ratio, the ratio of the condensate vapour partial pressure, $p_{\rm par}$ [dyne~cm$^{-2}$], to the saturation vapour pressure, $p_{\rm vap} (T)$ [dyne~cm$^{-2}$], $m_{0}$ [g] the mass of one molecular unit of the condensate and $n_{\rm v}$ [cm$^{-3}$] the number density of the condensing vapour.
The gaseous diffusion constant, $D$ [cm$^{2}$ s$^{-1}$], can be given by \citep{Jacobson2005}
\begin{equation}
    D = \frac{5}{16N_{\rm A}d_{i}^{2}\rho_{\rm a}}\sqrt{\frac{RT\bar{\mu}_{\rm a}}{2\pi}\left(\frac{\mu_{\rm v} + \bar{\mu}_{\rm a}}{\mu_{\rm v}}\right)},
\end{equation}
where $N_{\rm A}$ is Avogadro's number, $d_{i}$ [cm] the collision diameter, and $\mu_{v}$ the molecular weight of the condensable vapour.

In the free molecular regime, Kn $\gg$ 1, the condensation rate is given by \citep{Woitke2003}
\begin{equation}
\label{eq:cond_f}
    \left(\frac{dm}{dt}\right)_{\rm cond} = 4\pi r_{\rm c}^{2}v_{\rm th}m_{0} n_{\rm v} \left(1 - \frac{1}{S}\right),
\end{equation}
where $v_{\rm th}$ = $\sqrt{k_{\rm b}T/2\pi m_{v}}$ [cm s$^{-1}$] is the thermal velocity of the condensable vapour, with $m_{v}$ [g] the mass of the condensable vapour.
We note that we have ignored the effects of latent heat and the Kelvin effect in  Eqs. \ref{eq:cond_d} and \ref{eq:cond_f}, which can be modified to take into account these and additional kinetic effects near the surface of the grain \citep[e.g.][]{Toon1989, Lavvas2011, Gao2018b}.

\subsection{Homogeneous nucleation}

Homogeneous nucleation is the processes by which supersaturated vapour of a single gas species forms clusters of molecules which, in favourable thermochemical conditions, can then overcome the energy barrier of the nucleation process.
These clusters can then grow to form seed particle sized solid materials of $\sim$1 nm, and provide surfaces for condensable species to interact with.
This makes homogeneous nucleation an important consideration, as this process is responsible for forming the first surfaces, or cloud condensation nuclei (CCN), in the cloud formation system.

We follow the modified classical nucleation theory (MCNT) first outlined in \citet{Draine1977} and \citet{Gail1984}, examined in a sub-stellar atmosphere context in several studies \citep[e.g.][]{Helling2013, Lee2018, Sindel2022, Kiefer2023}.
The homogeneous nucleation rate, $J_{\rm hom}$ [cm$^{-3}$ s$^{-1}$], is given by \citep{Helling2013}
\begin{equation}
    J_{\rm hom} = \frac{\mathring{f}(1)}{\tau_{\rm gr}(1, N_{*})}Z(N_{*})\exp\left[(N_{*} - 1) \ln S(T) - \frac{\Delta G(N_{*})}{RT}\right],
\end{equation}
where $\mathring{f}(1)$ [cm$^{-3}$] is the equilibrium number density of the nucleating vapour and $\tau_{\rm gr}$(1, $N_{*}$) [s] the growth timescale of the critical cluster size $N_{*}$ given by 
\begin{equation}
    \tau_{\rm gr}^{-1} = 4\pi r_{0}^{2} N_{*}^{2/3} \alpha n_{\rm v} \sqrt{\frac{k_{\rm b}T}{2\pi m_{0}}},
\end{equation}
where r$_{0}$ [cm] is the vapour monomer radius, $\alpha$ an efficiency factor and n$_{\rm v}$ [cm$^{-1}$] the number density of the nucleating vapour.
The Gibbs free energy expression is given by
\begin{equation}
 \frac{\Delta G(N)}{RT} = \theta_{\infty} \frac{N-1}{ (N-1)^{1/3} + N_{\rm f}^{1/3}} \quad \mbox{with} \quad  \theta_{\infty} = \frac{4\pi r_{0}^{2} \sigma_{\infty}}{k_{\rm b}T},
 \label{eq:DGtheta}
\end{equation}
where $\sigma_{\infty}$(T) [erg cm$^{-2}$] is the bulk surface tension and $N_{\rm f}$ a fitting factor parameter.
The critical cluster size, $N_{*}$, is
\begin{equation}
 N_{*} - 1 = \frac{N_{*,\infty}}{8}\left[ 1 + \sqrt{1 + 2\left(\frac{N_{\rm f}}{N_{*,\infty}}\right)^{1/3}} - 2\left(\frac{N_{\rm f}}{N_{*,\infty}}\right)^{1/3}\right]^{3},
 \label{eq:N*}
\end{equation}
with
\begin{equation}
 N_{*,\infty} = \left(\frac{\frac{2}{3}\theta_{\infty}}{\ln S(T)}\right)^{3}.
\end{equation}
The Zeldovich factor, $Z$, is
\begin{equation}
 Z(N_{*}) = \left[\frac{\theta_{\infty}}{9 \pi(N_{*} - 1)^{4/3}}\frac{\left(1+2\left(\frac{N_{\rm f}}{N_{*} - 1}\right)^{1/3}\right)}{\left(1 + \left(\frac{N_{\rm f}}{N_{*} - 1}\right)^{1/3}\right)^{3}} \right]^{1/2}.
\end{equation}
For KCl homogeneous nucleation, we assume $\alpha$ = 1 and $N_{\rm f}$ = 5 following the analysis in \citet{Gail2013}, and the temperature dependent surface tension expression from \citet{Janz2013}.

\subsection{Seed particle evaporation}

Should all condensed species be evaporated from the surface of the particles, the seed particle core may evaporate back into the vapour phase.
If the average mass of the particles is within 0.1\% of the seed particle mass ($m_{\rm c}$ $<$ 1.001 $m_{\rm seed}$) and $dm/dt$ $<$ 0 we assume that the core is exposed to the atmosphere and undergoes evaporation.
The evaporation rate of seed particles, $J_{\rm evap}$ [cm$^{-3}$ s$^{-1}$], is approximated as
\begin{equation}
    J_{\rm evap} = -\frac{N_{\rm c}}{\tau_{\rm evap}},
\end{equation}
where $\tau_{\rm evap}$ is the seed particle evaporation timescale.
The evaporation rate of seed particles can be estimated as $\tau_{\rm evap}$ = {$m_{\rm seed}$/$\left|dm/dt\right|$ \citep{Lee2023}. 
However, this can prove numerically unstable, especially in strong evaporating regions. 
We therefore prefer a parameterised approach, adopting $\tau_{\rm evap}$ = 1 s in mini-cloud, which results in a smoother seed particle evaporation profile but still removing seed particles from regions of strong evaporation in one or two typical timesteps.

\subsection{Condensate vapour replenishment}

The atmosphere is continually refreshed with condensable vapour and excess vapour mixed deeper through connection with the interior, typically assumed to occur through deep convective motions.
To mimic this, at the bottom most layer only, we follow the relaxation approach from \citet{Komacek2022} given by
\begin{equation}
    \left(\frac{d\rho_{\rm v}}{dt}\right)_{\rm deep} = -\frac{\rho_{\rm v} - \rho_{\rm deep}}{\tau_{\rm deep}},
\end{equation}
where $\rho_{\rm deep}$ [g cm$^{-3}$] is the deep condensate vapour density and $\tau_{\rm deep}$ [s] the deep replenishment timescale.
The deep replenishment timescale can be set as a parameter or estimated from
\begin{equation}
    \tau_{\rm deep} = \frac{\alpha_{\rm mix} H_{\rm p, deep}^{2}}{K_{\rm zz, deep}},
\end{equation}
where $H_{\rm p, deep}$ [cm] is the atmospheric scale height, $K_{\rm zz, deep}$ [cm$^{2}$ s$^{-1}$] the eddy diffusion coefficient at the lowest layer respectively.
$\alpha_{\rm mix}$ is a scaling parameter that estimates the number of atmospheric scale heights of the mixing region in the deep atmosphere.

\section{Application to Y-dwarf KCl clouds}
\label{sec:GCM}

\begin{table*}
\centering
\caption{Adopted Exo-FMS GCM parameters for the test WISE 0359-54  Y-dwarf simulation. We use a cubed-sphere resolution of C48 ($\approx$ 192 $\times$ 96 in longitude $\times$ latitude).}
\begin{tabular}{c c c l}  \hline \hline
 Symbol & Value  & Unit & Description \\ \hline
 $T_{\rm int}$ &  458 & K & Internal temperature \\
 $p_{\rm 0}$ & 100 &  bar & Reference surface pressure \\
 $p_{\rm up}$ & 10$^{-4}$ &  bar & Upper boundary pressure \\
 $c_{\rm p}$ & 12045  &  J K$^{-1}$ kg$^{-1}$ & Specific heat capacity \\
 $R_{\rm d}$ & 3568 &  J K$^{-1}$ kg$^{-1}$  & Specific gas constant \\
 $\kappa$ &  0.296 & -  & Adiabatic coefficient \\
 $g_{\rm BD}$ & 288  & m s$^{-2}$ & Gravitational acceleration \\
 $R_{\rm BD}$ & 0.94  & R$_{\rm Jup}$ & Brown dwarf radius\\
 $P_{\rm BD}$ & 10 & hr & Rotational period \\
 $\left[{\rm M/H}\right]$ & 0 & - & $\log_{10}$ solar metallicity \\
 $p_{\rm amp}$ & 10 & bar & Perturbation boundary layer \\
 $T_{\rm amp}$ & 4.32 $\cdot$ 10$^{-5}$ & K s$^{-1}$ & Perturbation amplitude \\
 $\alpha$ & 1 & - & MLT scale parameter \\
 $\beta$ & 2.2 & - & Overshooting parameter \\
 $\tau_{\rm storm}$ & 10$^{5}$ & s & Storm timescale \\
 $\tau_{\rm drag}$ & 10$^{5}$ & s & Basal drag timescale \\
 $\Delta$ t$_{\rm hyd}$ & 60  & s & Hydrodynamic time-step \\
 $\Delta$ t$_{\rm rad}$  & 600 & s & Radiative time-step \\
 $\Delta$ t$_{\rm MLT}$ & 0.5  & s & Mixing length theory time-step \\ 
 $\Delta$ t$_{\rm ch}$  & 600 & s & mini-chem time-step \\
 $\Delta$ t$_{\rm cld}$  & 600 & s & mini-cloud time-step \\
 N$_{\rm v}$ & 60  & - & Vertical resolution \\
 d$_{\rm 4}$ & 0.16  & - & $\mathcal{O}$(4) divergence dampening coef. \\
\hline
\end{tabular}
\label{tab:GCM_parameters}
\end{table*}

In an initial test of our cloud scheme, we use the Exo-FMS GCM \citep[e.g.][]{Lee2021} following the simulation setup of \citet{Lee2024}, but replacing the \citet{Komacek2022} saturation adjustment cloud model with our microphysical 2-moment scheme.
All other physics and chemistry modules are kept as in \citet{Lee2024}, which includes mixing length theory (MLT), correlated-k radiative-transfer and the mini-chem kinetic chemistry scheme.
We assume a minimum $K_{\rm zz}$ = 10$^{5}$ cm$^{2}$ s$^{-1}$ diffusive mixing in the atmosphere, which is an estimate of the natural background minimal turbulent vertical mixing occurring in the radiative region of sub-stellar atmospheres \citep[e.g.][]{Ackerman2001, Christie2021}.
We model a WISE 0359-54 parameter regime setup, taking the values from \citet{Kothari2024} ($T_{\rm int}$ = 458 K, [M/H] = 0, log g = 4.46), who performed forward and retrieval modelling on the JWST data presented in \citet{Beiler2023}.
Following \citet{Showman2019}, the thermal perturbation amplitude at the radiative-convective boundary (RCB) is estimated to be $\sim$4.32$\cdot$10$^{-5}$ K s$^{-1}$.
For the thermal perturbation scheme, we assume the radiative-convective-boundary (RCB) is at 10 bar.
The rotational period of WISE 0359-54 is unknown, so we assume a Jupiter-like rotational period of 10 hours.

Our forcing and drag regime allows a quick statistical convergence of the atmospheric dynamics and composition, on par with the convergence rates for cooler objects simulated in \citet{Tan2021} and the $T_{\rm eff}$ = 500 K model in \citet{Lee2023b}.}
We perform the first 100 (Earth) days of simulation with the cloud component but without thermal feedback, after which the cloud opacity is ramped up slowly for the next 100 days. 
The simulation including the full cloud opacity in then performed for the final 900 days, for a total of 1100 days of simulation.
To examine the variability characteristics, we then run the model for an additional 96 hours, taking snapshots every hour of the simulation output.
We use the average values taken across this final 96 hours as the final results presented in the paper.

The chemical equilibrium deep volume mixing ratio of KCl is $\approx$10$^{-7}$ at [M/H] = 0 \citep{Woitke2018}, which we assume is the deep reservoir condensable vapour abundance.
We assume all Na has condensed in the atmosphere and do not include the opacity of gas phase Na and K in the GCM.
We assume an initial cloud and condensable vapour free atmosphere, with the lower boundary of the atmosphere assumed to be cloud free and replenished to the deep vapour reservoir value across each timestep.
We use the DLSODE\footnote{\url{https://computing.llnl.gov/projects/odepack/software}} stiff ordinary differential equation solver \citep{Radhakrishnan1993} to integrate the coupled ODE system inside the GCM.

We assume a pure KCl cloud composition with $\rho_{d}$ = 1.99 g cm$^{-3}$, $\mu_{v}$ = 74.55 g mol$^{-1}$, $r_{\rm seed}$ = 1 nm.
The vapour pressure expression for KCl is taken from \citet{Morley2012} and temperature dependent surface tension expression from \citet{Janz2013}.
Background gas conditions are self-consistently calculated in the GCM from the local T-p conditions and volume mixing ratios of gas phase species from the mini-chem kinetic chemistry module.
In the GCM, the moment values are converted to volume or mass mixing ratios through the relations 
\begin{equation}
    q_{0} = \frac{N_{\rm c}}{n_{\rm a}},
\end{equation}
\begin{equation}
    q_{1} = \frac{\rho_{\rm c}}{\rho_{\rm a}},
\end{equation}
and
\begin{equation}
    q_{\rm v} = \frac{\rho_{\rm v}}{\rho_{\rm a}},
\end{equation}
where $n_{\rm a}$ [cm$^{-3}$] is the atmospheric number density.
These mixing ratio tracers are the advected quantities inside the GCM dynamical core and other transport routines such as vertical settling and diffusion.

Eq. \ref{eq:vf} is used to calculate the settling velocities for the cloud particles inside the GCM.
We assume the settling velocity for all moments to occur at the mass weighted mean particle radius, $r_{\rm c}$, as given by Eq. \ref{eq:r_c}.
This approximation is valid when the size distribution can be approximated by the delta function, as adopted in the current 2-moment model. 
Strictly, each moment should have its own settling velocity \citep[e.g.][]{Woitke2020} as intuitively different parts of the particle size distribution, represented by the moment values, should settle at different rates (see also Appendix \ref{app:b}). 
However, this can lead to difficult to resolve numerical issues if a moment can settle faster into a layer where the other moments are negligible.

\subsection{Cloud opacity feedback}

To calculate the wavelength dependent cloud opacity, single scattering albedo and asymmetry factor, we follow a similar scheme to the cloud opacity calculations in \citet{Lee2023}.
For small size parameters (x $<$ 0.01) the Rayleigh limit expressions are used \citep{Bohren1983} and for large size parameters (x $>$ 10) Modified Anomalous Diffraction Theory (MADT) is used following \citet{Moosmuler2018}.
For intermediate size parameters, the LX-MIE code of \citet{Kitzmann2018} is applied.
We assume a well peaked, monodisperse distribution, where r$_{\rm c}$ is a representative size for the opacity of the size distribution, in line with the assumptions taken when constructing the moment scheme.
This can lead to inaccuracy when considering the full particle size distribution (e.g. Appendix \ref{app:dist}), as well as potentially being affected by Mie resonances \citep[e.g.][]{Hansen1974}, but is much faster to calculate and more amenable to large scale simulations such as GCMs, while still retaining the gross large scale effects of cloud opacity on the atmosphere.
Other schemes taking into account an assumed particle size distribution can be considered in future studies.

We take the real and imaginary refractive indices of KCl salt from \citet{Palik1985} and \citet{Querry1987}.
The MADT approach typically results in inaccuracies when the real refractive index is large (n $\gtrsim$ 1.5), which the real index of KCl reaches frequently across the wavelength regime of the database tables.
However, compared to the full Mie calculation, MADT offers much more computational efficiency for the large size parameter regime while retaining the salient effects of cloud opacity feedback, making it highly useful for large scale simulations such as GCMs.

For the asymmetry parameter, we assume $g$ = 0 for the Rayleigh regime, and take the value from LX-MIE for the intermediate size parameter regime.
However, MADT cannot directly calculate the asymmetry parameter, and so we follow the parametrisation from \citet{Ehlers2023}
\begin{equation}
    g = {\rm min}[C(m)x^{2},0.9],
\end{equation}
where $C(m)$ is a function of the complex refractive index \citep{Ehlers2023}. 
We limit $g$ to a maximum of 0.9 to attempt to capture the leveling off of the asymmetry parameter at large size parameters \citep[e.g.][]{Cuzzi2014} and avoid unphysical values of $g$.

\section{WISE 0359-54 GCM results}

\begin{figure*}
    \centering
    \includegraphics[width=0.49\linewidth]{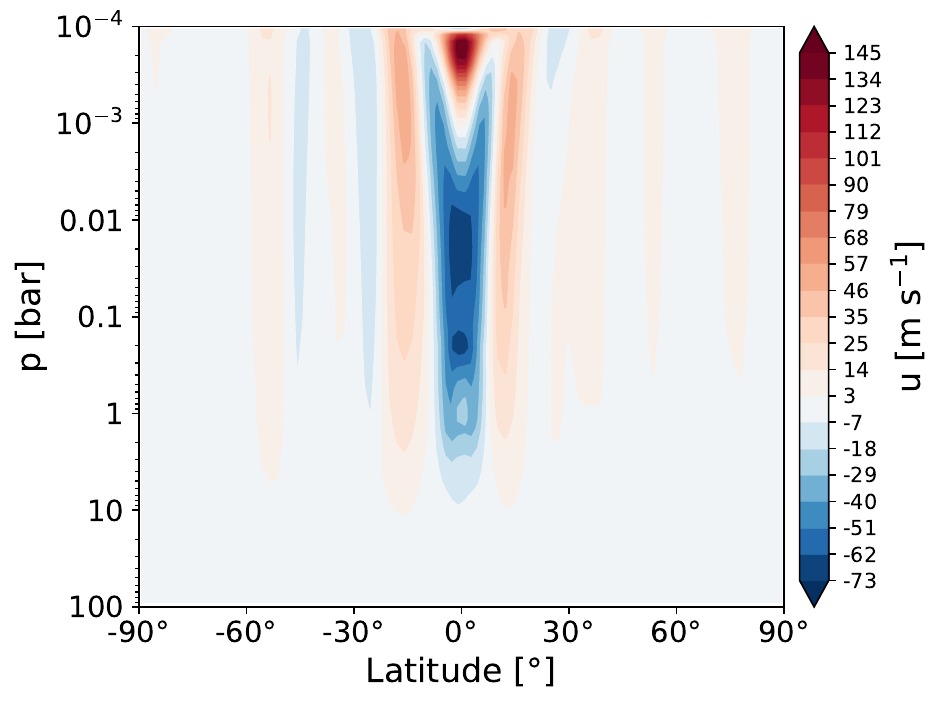}
    \includegraphics[width=0.49\linewidth]{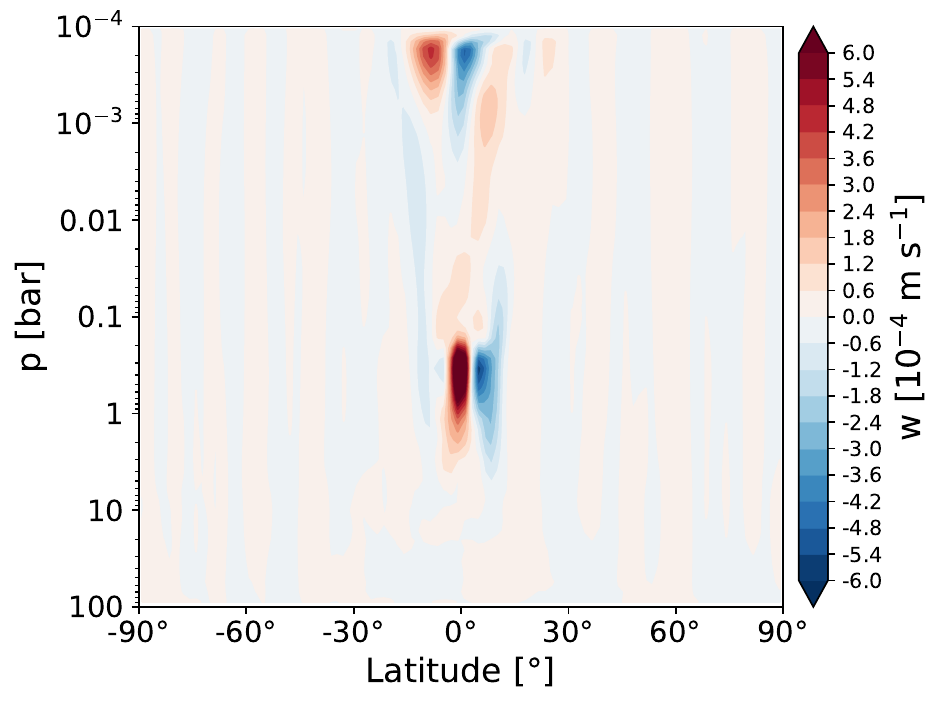}
    \caption{Dynamical properties of the atmosphere after 1104 days of simulation, averaged across the final 4 days. 
    Left: zonal mean zonal velocity. 
    Right: zonal mean vertical velocity.
    Little dynamical structure is present outside the equatorial region, leaving a generally sluggish and stagnant atmosphere.}
    \label{fig:zonal}
\end{figure*}

\begin{figure*}
    \centering
    \includegraphics[width=0.47\linewidth]{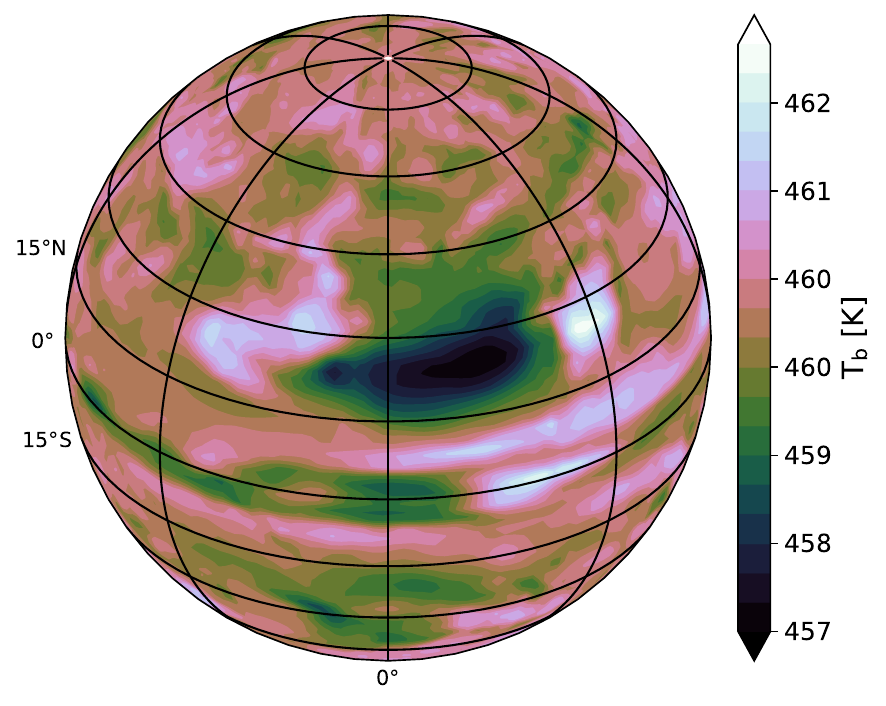}
    \includegraphics[width=0.47\linewidth]{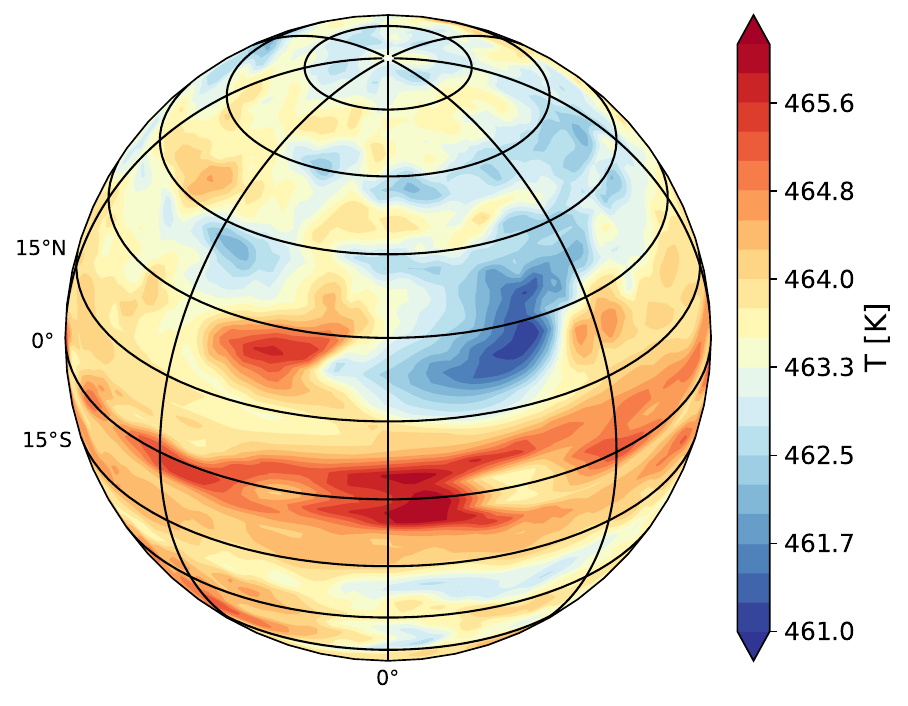}
    \includegraphics[width=0.47\linewidth]{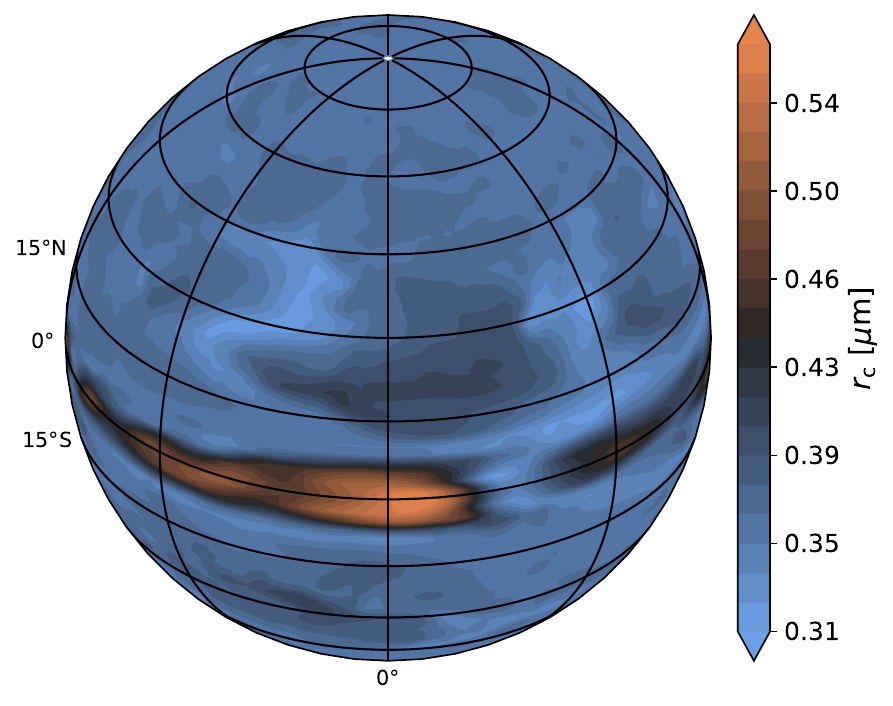}
    \includegraphics[width=0.47\linewidth]{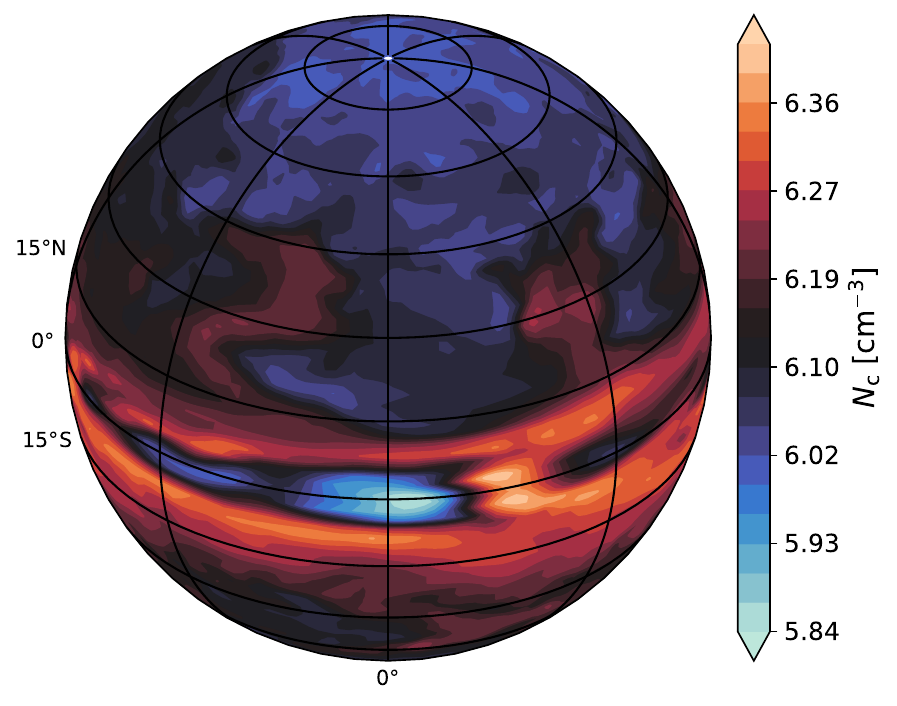}
    \includegraphics[width=0.47\linewidth]{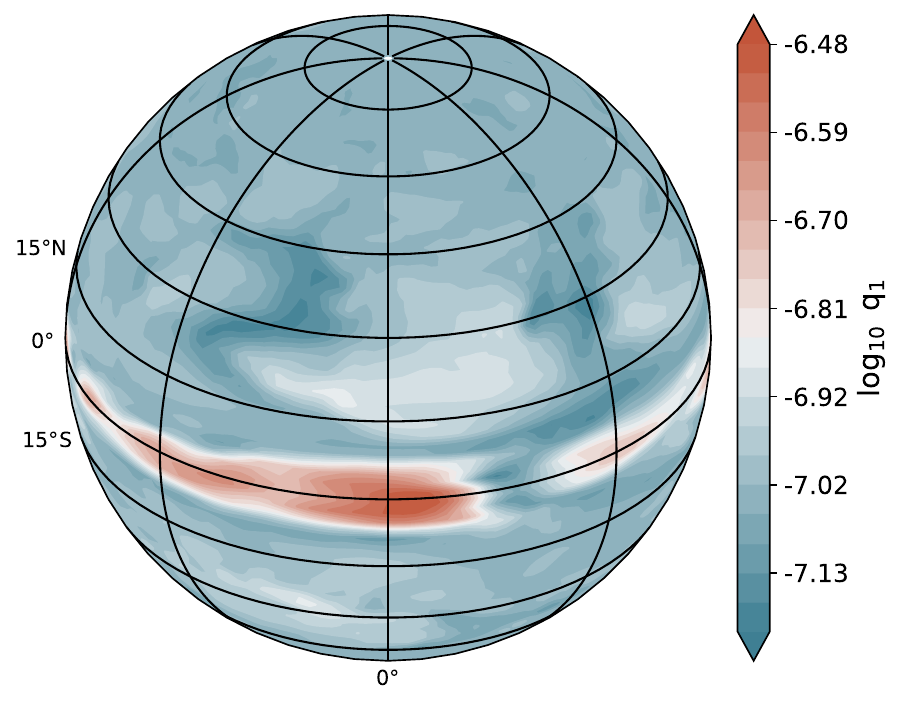}
    \includegraphics[width=0.47\linewidth]{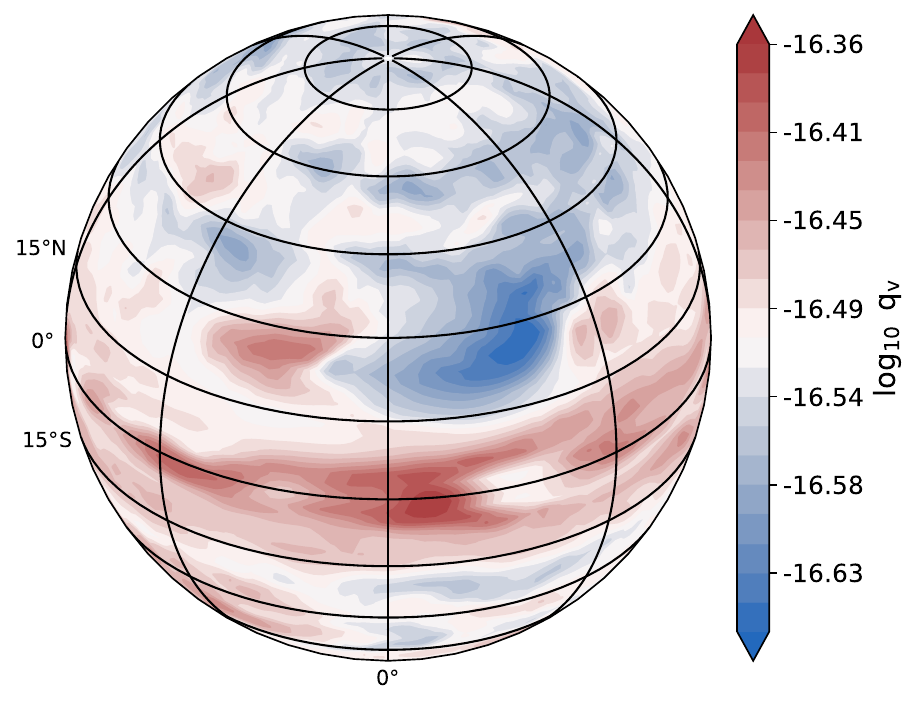}
    \caption{Latitude-longitude projections of the atmospheric conditions at 1104 days of simulation, averaged across the final 4 days at the 0.4 bar pressure level. 
    Top left: brightness temperature, $T_{\rm b}$ [K]. 
    Top right: atmospheric temperature. 
    Middle left: mass weighted average cloud particle size. 
    Middle right: number density of the cloud particles.
    Bottom left: Mixing ratio of the first moment ($q_{1}$).
    Bottom right: Mixing ratio of the condensate vapour ($q_{\rm v}$).
    }
    \label{fig:GCM_maps}
\end{figure*}

\begin{figure*}
    \centering
    \includegraphics[width=0.49\linewidth]{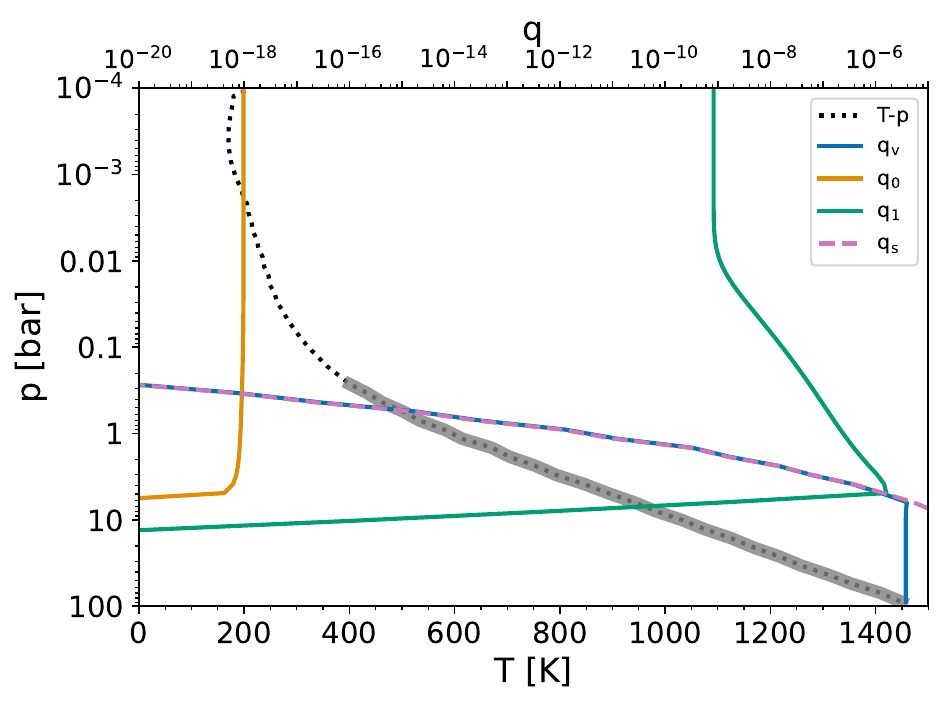}
    \includegraphics[width=0.49\linewidth]{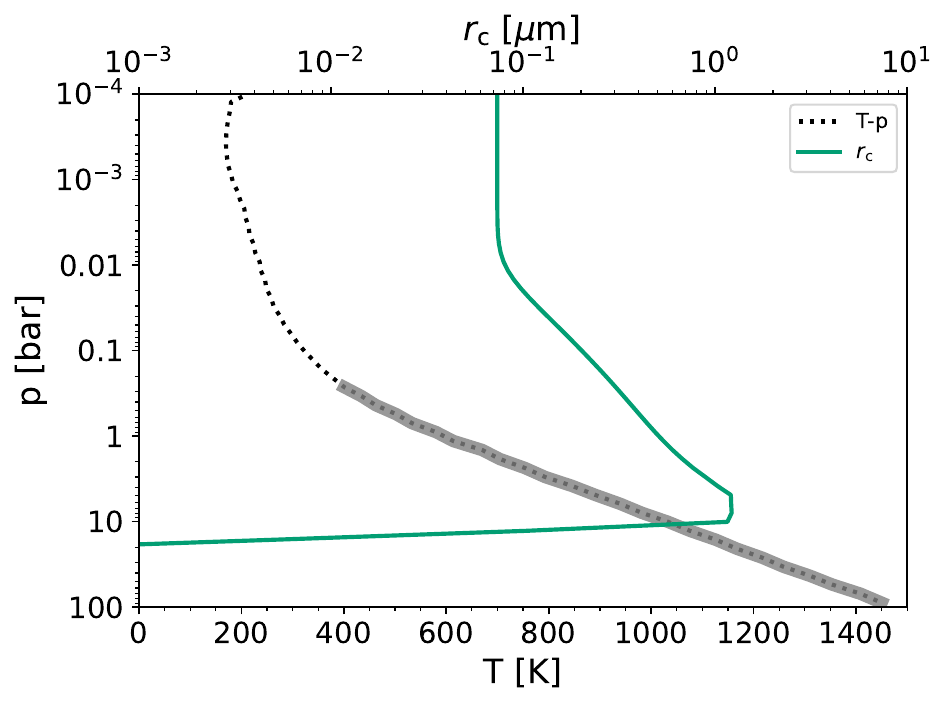}
    \includegraphics[width=0.49\linewidth]{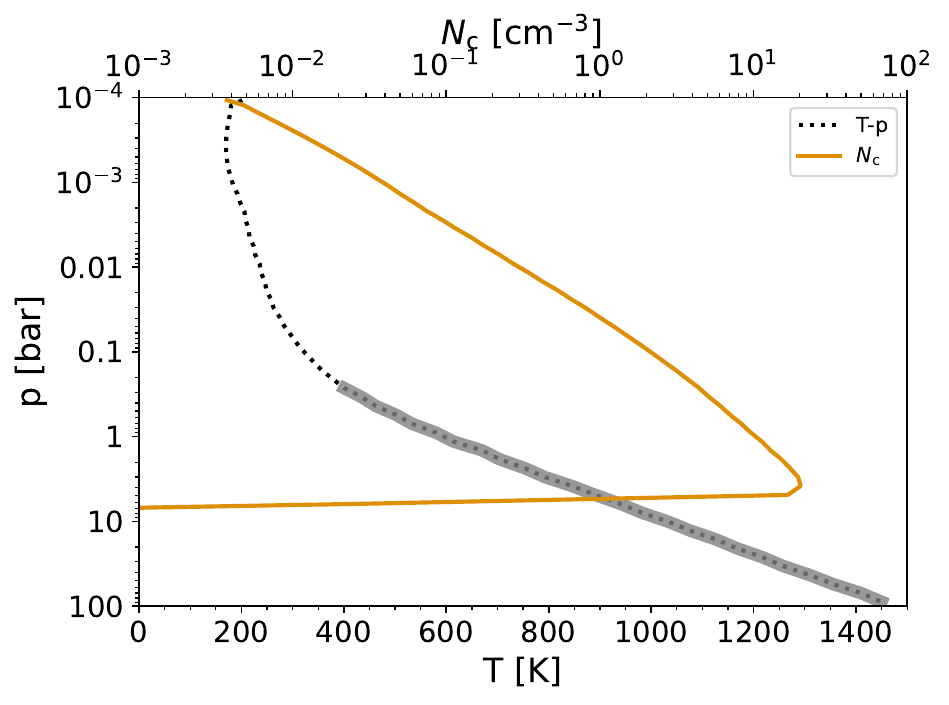}
    \includegraphics[width=0.49\linewidth]{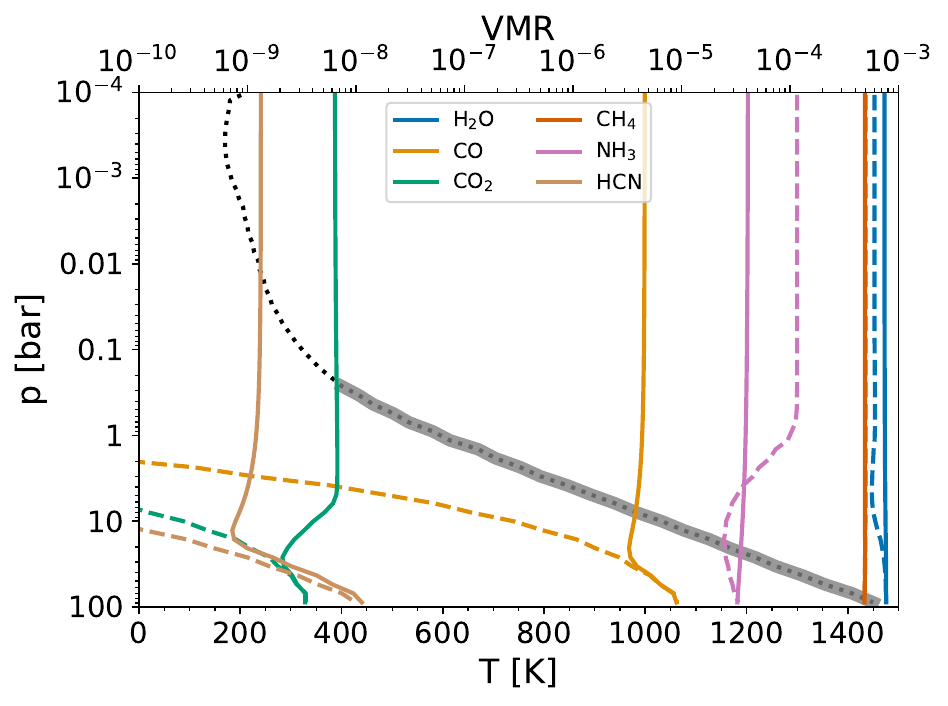}
    \caption{Globally averaged properties of the GCM after 1104 days of simulation, averaged across the final 4 days of simulation. 
    The dotted line shows the globally averaged T-p profile, with the grey shading denoting the approximate convective region of the atmosphere.
    Top left: mixing ratio values for the moments $q_{0}$ and $q_{1}$ and saturation mixing ratio, $q_{\rm s}$ (pink dashed line).
    Top right: mass weighted mean cloud particle size, $r_{\rm c}$ [$\mu$m].
    Bottom left: number density of the cloud particles, $N_{\rm c}$ [cm$^{-3}$].
    Bottom right: volume mixing ratio (VMR) of the main chemical species in the mini-chem kinetic chemistry scheme (solid lines) and chemical equilibrium abundances (dashed lines).
    }
    \label{fig:vert_av}
\end{figure*}

In this section, we present the results of the WISE 0359-54 GCM simulation coupled to the 2-moment cloud microphysical model.
Figure \ref{fig:zonal} presents the zonal mean zonal velocity and zonal mean vertical velocity averaged from the final 4 days of simulation.
This shows similar zonal jet structures to previous brown dwarf modeling using a similar drag and forcing parameter set \citep{Tan2021}. 
The dynamical features are dominated by the counter rotating jet at the equator, flanked by similar strength pro-rotating jets. 
Mid latitudes show multiple jet features, but with significantly weaker zonal velocity, suggesting slower dynamical evolution at higher latitudes.
For the vertical velocities, the GCM exhibits very weak vertical transport with maximum velocities on the order of $\sim$6 $\cdot$ 10$^{-4}$ m s$^{-1}$. 
Vertical mixing is therefore dominated by the prescribed minimal $K_{\rm zz}$ = 10$^{-5}$ cm$^{2}$ s$^{-1}$, suggesting vertical motions being convective and turbulence driven, rather than any global scale vertical motion system.
We find that the MLT scheme does not produce values above this minimum $K_{\rm zz}$ value, suggesting only weak convective motions are required to retain the deep adiabatic gradient.

Figure \ref{fig:GCM_maps} presents projections of the mean output of the final 4 days of simulation. 
The brightness temperature plot shows patchy structures with variations of $\sim$4 K of the internal temperature of the simulation.
This patchiness is long-lived in the GCM simulations, with the features in the simulations lasting several rotation rates.
For example, the large cooler region in Figure \ref{fig:GCM_maps} is present all through the final 4 days of the simulation.
Slow dynamical evolution at equatorial regions are seen, with features moving across the face of brown dwarf for many days, a product of the faster zonal velocities occurring at the jet region there.
This is also seen in the temperature map, where cooler and warmer patches varying by $\approx$4 K are seen at the 0.4 bar pressure level, near the photosphere of the brown dwarf.

This patchiness is also seen in the cloud structures on a global scale in Figure \ref{fig:GCM_maps}.
The equatorial regions contain generally larger particles compared to the higher latitudes, though only by around $\approx$0.1-0.2 $\mu$m at the 0.4 bar pressure level.
The equator region is also patchy, with the cloud particle size varying with longitude, suggesting that zonal velocities are not strong enough to horizontally homogenise the cloud structure there.
The cloud particle number density is also variable across the globe. 
Most concentration of cloud particles are near the equatorial regions, through regions of depletion are present at the equatorial latitudes.
However, the overall variations are small, with the 0.4 bar level only showing variations of $\approx$0.5 cm$^{-3}$ across the globe.

Figure \ref{fig:GCM_maps} also shows the first moment and condensate vapour mixing ratio at the same 0.4 bar pressure level. 
From these results we suggest that slight variation in the vapour distribution as well as sensitivity of the condensation rate to temperature variations are the cause of the variation in the cloud properties across the globe. 
We suggest that the slight variations in temperature as well as vapour from dynamical motions at the patchy and equatorial regions affect the end result of the cloud microphysics.
However, the dynamical timescales are much slower in the Y-dwarf objects compared to the hotter, more strongly driven L and T dwarf simulations performed in \citet{Tan2021} and \citet{Lee2023b} allowing the patchy features to persist over many rotational periods.
However, we not that these slight variations are still very minor compared to more strongly driven brown dwarfs \citep[e.g.][]{Lee2024} resulting in comparably a more homogeneous cloud structure.

Figure \ref{fig:vert_av} presents the vertical global averaged values for the cloud and chemical species characteristics across the final 4 days of the simulation.
This shows that the cloud structure extends to the top of the atmosphere, but varies with particle size and number density with height.
The maximum mean particle size at the cloud base is $\sim$1.5 $\mu$m at around the 3 bar pressure level. 
The particle size quickly drops to below sub-micron sizes, with particles of size $\sim$0.07 $\mu$m able to remain lofted up to the top of the atmosphere.
The number density of the cloud particles shows a typical structure, with a drop-off of several orders of magnitude with height starting from the cloud base.

Figure \ref{fig:vert_av} also shows the vertical profiles of the main chemical species in the atmosphere produced in the GCM.
The main chemical species in the atmosphere are deeply quenched, below the 10 bar level, similar to that seen in other brown dwarf kinetic models \citep[e.g.][]{Zahnle2014}.
This leads to a substantial CO abundance in the atmosphere compared to the expected composition assuming chemical equilibrium.
Our results suggest the main chemical signatures in the atmosphere should be CO, \ce{NH3}, \ce{CH4} and \ce{H2O}.
The quenched mixing ratios of these species align well with the median retrieved values found in \citet{Kothari2024} for WISE 0359-54, suggesting the GCM is decently capturing the mixing properties of the atmosphere.

\section{Emission spectra and variability characteristics}
\label{sec:postpro}

\begin{figure*}
    \centering
    \includegraphics[width=0.95\linewidth]{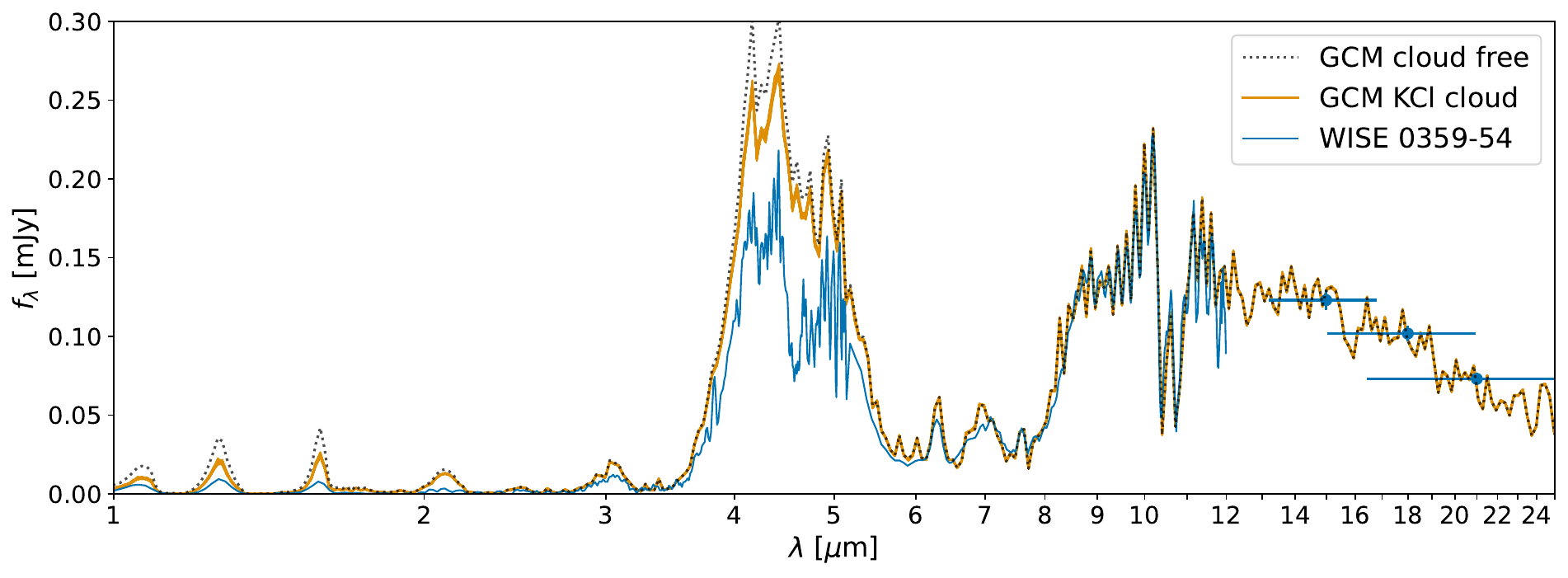}
    \caption{Comparison between the post-processed spectra from the GCM and the WISE 0359-54 JWST spectral data from \citet{Beiler2023}.
    The width of the orange line shows the extent of the maximum and minimum flux across the last 96 hours of simulation.
    The dotted black line shows the resultant spectra assuming no cloud opacity.
    The GCM shows decent agreement beyond 7 $\mu$m but is not too consistent with the data at bluer wavelengths, suggesting additional cloud opacity is required to dampen the spectral features at the bluer wavelengths.}
    \label{fig:obs_comp}
\end{figure*}

\begin{figure}
    \centering
    \includegraphics[width=0.95\linewidth]{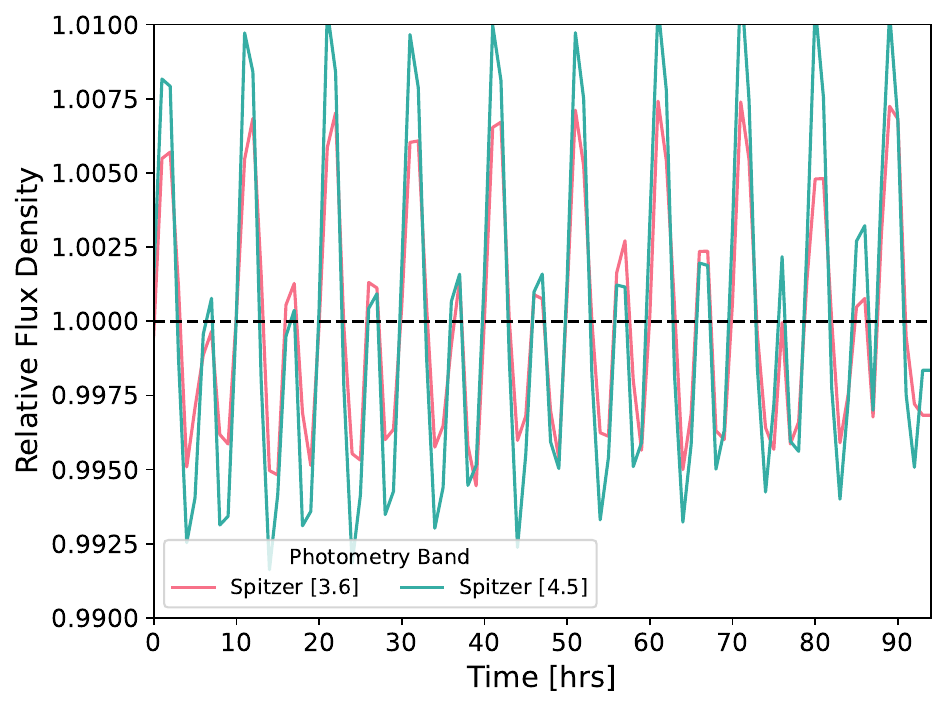}
    \caption{Relative flux density for the Spitzer [3.6] and [4.5] bands produced from the last 96 hours of the GCM simulation, at an inclination of 90$^{\circ}$. 
    This shows a 10-hour peak-to-peak rotational component with $\sim$0.5-1 \% variability. 
    Small scale inter-rotational periodic variability is also seen in the model.}
    \label{fig:obs_var}
\end{figure}

We use the 3D radiative-transfer code, gCMCRT \citep{Lee2022}, to post-process the GCM simulation results.
We assume an exponential particle size distribution, taking the mean values from the GCM moment results to reconstruct the particle size distribution in each cell (Appendix \ref{app:dist}).

In Figure \ref{fig:obs_comp} we compare the emission spectra post-processes from the GCM to the JWST data in \citet{Beiler2023}, who presented a wide wavelength JWST spectra of the Y-dwarf WISE J035934.06-540154.6, finding that models with weak mixing $K_{\rm zz}$ $\sim$ 10$^{4}$ cm$^{2}$ s$^{-1}$ best reproduced the JWST spectral data.
However, \citet{Kothari2024} performed retrieval and forward model analysis of the \citet{Beiler2023} data, finding that strong mixing of $K_{\rm zz}$ $\sim$ 10$^{9}$ cm$^{2}$ s$^{-1}$ was required to best fit the data.
Our study shows that the \ce{NH3} feature is well fit by the GCM simulation, with it's minimal $K_{\rm zz}$ $\sim$ 10$^{5}$ cm$^{2}$ s$^{-1}$, suggesting weak vertical mixing is enough to quench chemical species to their observed mixing ratios.

However, for wavelengths bluer than $\sim$7 $\mu$m, the GCM generally over-predicts the flux compared to the observations, with the KCl cloud not able to reduce flux in the opacity window regions effectively.
The minor impact of the KCl cloud is not surprising, as the abundance of KCl vapors is insufficient to form an optically thick cloud for solar composition atmospheres \citep{Fortney05}.
We therefore suggest that additional cloud species are required to increase the cloud opacity to reduce the flux to the observed levels.
\ce{Na2S} and ZnS, or NaCl should \ce{Na2S} formation be unfavorable due to \ce{Na2S}'s high surface tension \citep[e.g.][]{Powell2024}, would provide more cloud mass in this temperature regime \citep{Morley2012}.
Depending on the size of \ce{Na2S} particles, they may be able to efficiently mix upward from its deeper condensation pressure than KCl.
ZnS and NaCl are expected to form at similar pressures and temperatures to KCl at solar metallicity \citep{Morley2012}.
Considering these additional cloud species will be subject to future investigations and an update to the theory to include mixed material grains such as those considered in \citet{Helling2008}. 

Another possibility to increase variability is the formation of \ce{H2O} clouds high in the atmosphere for this Y-dwarf temperature regime \citep[e.g.][]{Morley2012}.
Our models suggest that \ce{KCl} particles may be a suitable nucleation site for the condensation of \ce{H2O} as suggested in \citet{Lee2018}, as without condensation sites the required supersaturation for \ce{H2O} self-nucleation would be high, requiring lower temperatures in the upper atmosphere than those simulated here.
This scenario has been modeled in detail in \citet{Mang2022} who used a detailed microphysical model to examine \ce{KCl} condensation sites for \ce{H2O} as well as \ce{H2O} self-nucleation in the Y-dwarf regime.
However, the methodology presented in this paper is unable to capture the condensation of \ce{H2O}, which we will address in future iterations of the scheme.

Figure \ref{fig:obs_var} shows the Spitzer [3.6] and [4.5] bands relative flux curves from the GCM results.
This shows a 10-hour peak-to-peak rotational component in the atmosphere, with a 0.5-1\% variation, lower than the observed Y-dwarf regime Spitzer variability from the \citet{Cushing2016} and \citet{Leggett2016} observations.
This suggests the GCM is capturing the periodic variability characteristics of the Y-dwarf population in this regime, but not the magnitude of the variability. 
Additional cloud opacity with thermal feedback or stronger forcing parameters are possible ways to increase the variability amplitude in the GCM simulation.
Our simulation also shows inter-periodic variability, suggesting additional smaller-scale, long-term persistent patchiness is present, which correlates to the GCM results (Figure \ref{fig:GCM_maps}).

\section{Discussion}
\label{sec:disc}

To estimate the dynamical regime of our GCM simulation, we compare qualitatively with the shallow water models performed in \citet{Hammond2023}.
The thermal timescale is approximately given by \citep[e.g.][]{Zhang2014}
\begin{equation}
    \tau_{\rm rad} = \frac{p}{g}\frac{c_{\rm p}}{4\sigma T_{\rm eff}^{3}},
\end{equation}
with the non-dimensional radiative timescale given by $\Gamma$ = 2$\Omega$$\tau_{\rm rad}$.
Assuming p $\approx$ 4 $\cdot$ 10$^{4}$ Pa and $T_{\rm eff}$ = 548 K, this gives  $\tau_{\rm rad}$ $\approx$  4.5 $\cdot$ 10$^{4}$ s and $\Gamma$ $\approx$ 16 for our simulation.
Following \citet{Hammond2023}, the thermal Rossby number for our simulation is Ro$_{\rm T}$ $\approx$ 10$^{-3}$.
This places our simulation in the bottom left corner of the parameter space explored by \citet{Hammond2023} (their Figure 1.), which conforms to the general dynamical patterns found in our simulation of weak or no global jet or vortex features, with a stagnant atmosphere outside the equatorial region.
However, weak jets are found near the equator where cloud coverage is thickest in our simulation, just above the cloud base (Fig. \ref{fig:zonal}), suggesting that the KCl clouds are able to adequately force the appearance of weak dynamical jet features in the atmosphere.
We suggest that these induced dynamical features would then be primarily responsible for the majority of the non-rotational component variability seen in these objects for this dynamical regime.

The current study does not include mixed or dirty grain compositions \citep[e.g.][]{Helling2008, Woitke2020}, making the current model limited to single composition systems such as KCl.
This can be included through modifying the moment equations in a similar manner to that considered in \citet{Helling2008} to take into account the condensation rate of different materials onto the mixed grain surface.
Mixed grain material in our mass moment framework will be examined in a follow up study.

In this study, we assumed a single particle particle size for deriving the settling velocity of the moments as well as the cloud opacity.
This simplification enables efficient calculation, important for GCM simulations, but accuracy can be improved through assuming a particle size distribution derived from the values of the moments which may be more appropriate for detailed 1D modelling efforts using this method.
In Appendix \ref{app:dist} we describe some relations using the moment results and deriving moment dependent vertical velocities and diffusion coupling, as well as integrating the particle size distribution to calculate opacities.

\section{Conclusions}
\label{sec:conc}

In this study, we present a 2-moment cloud bulk microphysical scheme for sub-stellar atmospheres that can be used for general time-dependent atmosphere simulations such as GCMs.
Overall, our current model follows closely and combines the methods in \citet{Ohno2018} and \citet{Lee2023} into a unified generalised bulk microphysical approach, combining homogeneous nucleation, condensation and collisional processes together.

In a test case, we simulated KCl cloud formation for a Y-dwarf WISE 0359-54 parameter test case (\citet{Kothari2024}; $T_{\rm int}$ = 458 K, [M/H] = 0, log g = 4.46, $P_{\rm rot}$ = 10 hr) using the 3D Exo-FMS GCM coupled to the new 2-moment scheme.
The GCM produces a generally sluggish and stagnant atmosphere, with near zero zonal and vertical velocities outside the equatorial region.
Compared to the weak jets expected from shallow water modelling \citep{Hammond2023}, our results suggest that the small thermal feedback from KCl clouds can help force the production of weak equatorial jets, on the order of Jupiter's equatorial region ($\sim$100 m s$^{-1}$).
Our GCM simulations show significant progress in modelling cold 3D brown dwarf atmospheres, including consistent chemistry, cloud microphysics and radiative-transfer, but several challenges and improvements with the current study remain to be resolved, with the most obvious being that more cloud opacity from KCl, or additional cloud species not considered in this study, is required to better fit the WISE 0359-54 JWST spectrum from \citep{Beiler2023}.
More complex GCM studies in this parameter regime are warranted to understand the mechanisms of Y-dwarf variability more clearly.

Overall, the current 2- moment cloud microphysical scheme is highly suitable and computationally efficient enough for inclusion in large scale GCM simulations.
The model can be extended to include `split' moments, typically called `cloud' and `rain' components in Earth science literature \citep[e.g.][]{Seifert2001} similar to \citet{Ohno2017} as well as include mixed composition grains such as those considered in \citet{Helling2008}.
Latent heat, heterogeneous nucleation and surface Kelvin effects will also be explored in future iterations.
The 2-moment schemes, as well as an implementation of the \citet{Komacek2022} saturation adjustment scheme are available online\footnote{\url{https://github.com/ELeeAstro/mini_cloud}}.

\begin{acknowledgements}
E.K.H. Lee is supported by the CSH through the Bernoulli Fellowship.
K.Ohno is supported by the JSPS KAKENHI Grant Number JP23K19072.
This work benefited from the 2024 Exoplanet Summer Program in the Other Worlds Laboratory (OWL) at the University of California, Santa Cruz, a program funded by the Heising-Simons Foundation and NASA.
\end{acknowledgements}

\bibliographystyle{aa} 
\bibliography{bib.bib} 

\begin{appendix}

\section{Full derivation of the moment equations}
\label{app:b}

    In general, one can describe the evolution of particle size distribution as
    \begin{equation}
    \begin{split}
        \frac{\partial f(m)}{\partial t}=&-\nabla\cdot(\mathbf{V}_{\rm wind}f(m))-\frac{\partial [v_{f}(m)f(m)]}{\partial z}-\frac{\partial}{\partial m}\left[\left(\frac{dm}{dt}\right)_{\rm cond}f(m)\right]\\
        & +\frac{1}{2}\int^{\rm m-m'}_{\rm 0}K(m',m-m')f(m')f(m-m')dm' \\
        & -f(m)\int^{\rm \infty}_{\rm 0}K(m,m)f(m')dm' + J(m),
    \end{split}
    \end{equation}
where $\mathbf{V}_{\rm wind}$ is the 3D velocity vector of atmospheric wind, $J(m)$ is the nucleation rate of particles with a mass of $m$.
In the right side, the first and the second terms stand for spatial advection due to atmospheric flow and particle's gravitational settling, the third term stands for the advection in a mass space driven by condensation or evaporation, the fourth and the fifth terms express the gain and loss of cloud particles through collisional sticking, and the sixth term expresses the new particle formation through nucleation.
Multiplying $m^{k}$ and integrating the equation over whole mass space, we obtain
\begin{equation}\label{eq:master_mom}
\begin{split}
    \frac{\partial M^{(k)}}{\partial t}=&-\nabla\cdot(\mathbf{V}_{\rm wind}M^{(k)})- \frac{\partial (\overline{v}_{\rm f}^{(k)}M^{(k)})}{\partial z}\\    
    &+ m_{\rm seed}^{k}J_{\rm hom} +\int_{\rm 0}^{\rm \infty}km^{k-1}\left( \frac{dm}{dt}\right)_{\rm cond}f(m)dm\\
    & +\frac{1}{2}\int^{\rm \infty}_{\rm 0}\int^{\rm \infty}_{\rm 0}[(m+m')^{k}-m^k-m'^{k}]\\
    &~~~~~~~~~~~~~~~~~~~~~~~\times 
    K(m',m)f(m')f(m)dmdm',
    \end{split}
    \end{equation}
where we have used the conversion of collision term given by Equation \eqref{eq:coll_mom} and assumed that $(dm/dt)f(m)$ converges to zero at $m\rightarrow0$ and $m\rightarrow\infty$.
The averaged terminal velocity $\overline{v}_{\rm f}$ is defined as
\begin{equation}
    \overline{v}_{\rm f}^{(k)}=\frac{\int_{\rm 0}^{\infty} m^{k}v_{f}(m)f(m)dm}{\int_{\rm 0}^{\infty} m^kf(m)dm}.
\end{equation}
We also express the homogeneous nucleation rate profile as $J(m)=J_{\rm hom}\delta(m-m_{\rm seed})$, where $\delta(x)$ is the delta function.

For the zeroth ($k=0$) and first moment ($k=1$), one can simplify Equation \eqref{eq:master_mom} as
\begin{equation}
\begin{split}
    \frac{\partial M^{(0)}}{\partial t}=&-\nabla\cdot(\mathbf{V}_{\rm wind}M^{(0)})- \frac{\partial (\overline{v}_{\rm f}^{(0)}M^{(0)})}{\partial z}\\    
    &+ J_{\rm hom} -\frac{1}{2}\int^{\rm \infty}_{\rm 0}\int^{\rm \infty}_{\rm 0}K(m',m)f(m')f(m)dmdm',
\end{split}
\end{equation}

\begin{equation}
\begin{split}
    \frac{\partial M^{(1)}}{\partial t}=&-\nabla\cdot(\mathbf{V}_{\rm wind}M^{(1)})- \frac{\partial (\overline{v}_{\rm f}^{(1)}M^{(1)})}{\partial z}\\    
    &+ m_{\rm seed}J_{\rm hom} +\int_{\rm 0}^{\rm \infty}\left( \frac{dm}{dt}\right)_{\rm cond}f(m)dm.
\end{split}
\end{equation}
These simplification reflect the fact that condensation does not change the total particle number ($N_{\rm c}=M^{(0)}$), whereas collisions do not change the total particle mass ($\rho_{\rm c}=M^{(1)}$) regardless of the expression of the size distribution, the condensation rate, and the collision kernel. 

\subsection{2-moment model}

To move forward from Equation \eqref{eq:master_mom}, one needs to assume a certain shape of the size distribution $f(m)$ to obtain explicit expressions of each source term.
The simplest size distribution is the delta function $f(m)=N_{\rm c}\delta(m-m_{\rm c})$, meaning that cloud particles at each vertical layer are characterized by a single particle size (mass).
This approach has been widely used in the models of grain growth in protoplanetary disks \citep[e.g.,][]{Sato2016}.
Since the assumed size distribution contains 2 unknowns, $N_{\rm c}$ and $m_{\rm c}$, two moments are needed to close the system.
Each moment is related to $N_{\rm c}$ and $m_{\rm c}$ as
\begin{equation}
    M^{(0)}=N_{\rm c}
\end{equation}
\begin{equation}
    M^{(1)}=\rho_{\rm c}=m_{\rm c}N_{\rm c}
\end{equation}

Inserting $f(m)=N_{\rm c}\delta(m-m_{\rm c})$ into Equation \eqref{eq:master_mom}, we obtain the master equation of the 2-moment model:
\begin{equation}
    \frac{\partial N_{\rm c}}{\partial t}=-\nabla\cdot(\mathbf{V}_{\rm wind}N_{\rm c})- \frac{\partial (v_{\rm f}(m_{\rm c})N_{\rm c})}{\partial z}+ J_{\rm hom} -\frac{1}{2}K(m_{\rm c},m_{\rm c})N_{\rm c}^2,
\end{equation}

\begin{equation}
    \frac{\partial \rho_{\rm c}}{\partial t}=-\nabla\cdot(\mathbf{V}_{\rm wind}\rho_{\rm c})- \frac{\partial (v_{\rm f}(m_{\rm c})\rho_{\rm c})}{\partial z}+ m_{\rm seed}J_{\rm hom} +N_{\rm c}\left( \frac{dm}{dt}\right)_{\rm cond,m=m_{\rm c}}.
\end{equation}
These expressions are equivalent to the model of \citet{Ohno2018}, though they approximated the wind transport by eddy diffusion for 1D framework and omitted homogeneous nucleation.
We note that the approximation of $f(m)=N_{\rm c}\delta(m-m_{\rm c})$ leads to the same moment-averaged terminal velocity for zeroth and first moment, i.e., $\overline{v}_{\rm f}^{(0)}=\overline{v}_{\rm f}^{(1)}=v_{\rm f}(m_{\rm c})$.

The 2-moment framework has been extensively utilized in the microphysical model of Earth clouds \citep[for review, see e.g.,][]{Morrison+20_Earth_cloud_review}.
We note that 2 moment framework does not necessary need to assume the delta function for the size distribution.
For instance, 2-moment models for Earth cloud often assume a gamma distribution \citep[e.g.,][]{Ziegler85,Schoenberg94}, given by
\begin{equation}\label{eq:gamma}
    f(m)=\frac{N_{\rm c}\beta^{\alpha+1}}{\Gamma(\alpha+1)} m^{\alpha}\exp(-\beta m),
\end{equation}
where $\Gamma(x)\equiv \int_{\rm 0}^{\infty} y^{x-1}e^{-y}dy$ is the gamma function.
The gamma distribution includes 3 unknown parameters ($N_{\rm c}$, $\alpha$, $\beta$), which require three moments to close the system. 
However, one could reduce the number of unknowns by treating one of them as a free parameter, which allows us to use the gamma distribution for 2-moment framework  \citep[$\alpha$ is typically fixed in Earth cloud models,][]{Ziegler85,Schoenberg94}. 
This approach can be useful if existing observations have already put constraint on the shape of size distribution, as in the cases of Earth clouds.
On the other hand, the shape of particle size distributions remain highly uncertain for brown dwarf and exoplanetary atmospheres.
Thus, we decided to keep adopting the simple delta function derivation for the present 2-moment model, as in \citet{Ohno2018}.

    \section{Size distribution effects}
    \label{app:dist}

    In this Appendix, we detail fitting particle mass and size distributions from the results of the moment equations. 
    We also show distribution integrated cloud opacities using the fit distributions.

    \subsection{Fitting mass and size distributions}

    \begin{figure*}
        \centering
        \includegraphics[width=0.45\linewidth]{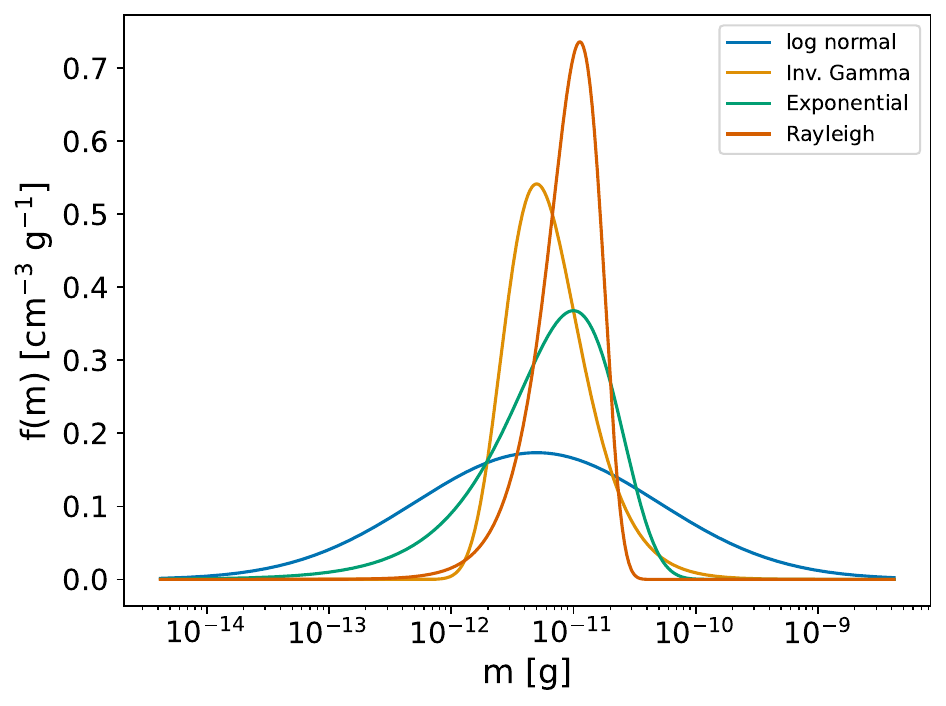}
        \includegraphics[width=0.45\linewidth]{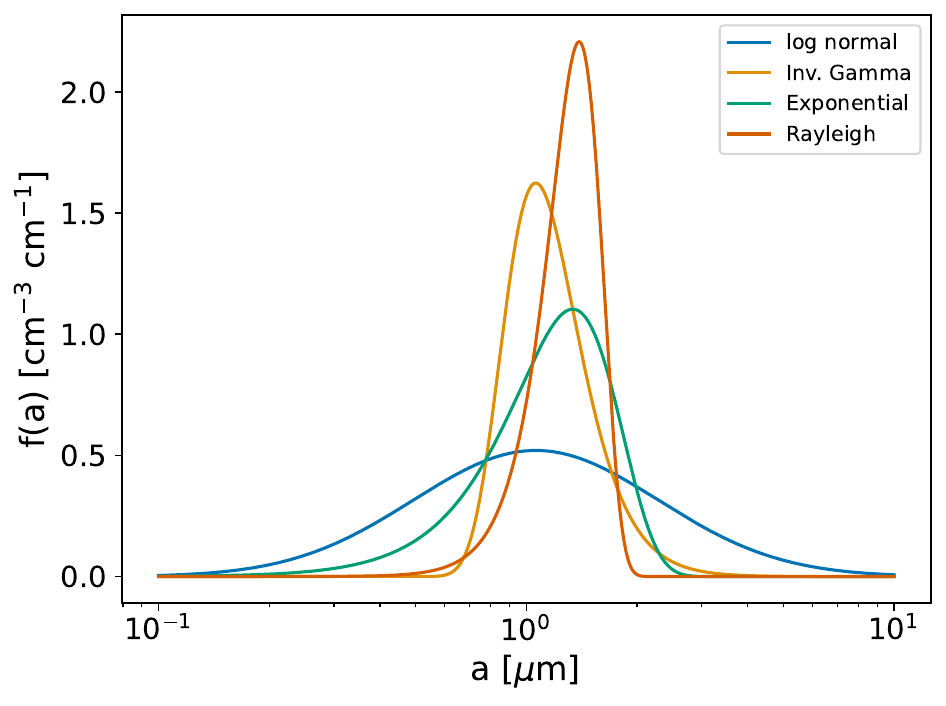}
        \caption{Particle mass distribution (left) and equivalent particle radius distribution (right).
        The difference in magnitude scale between the two is large, with the mass distributed over many more magnitudes compared to the radius.}
        \label{fig:dist}
    \end{figure*}
        
    Deriving continuous size distributions from the moment solutions is first performed by deriving the expectation value $E[m] [g]$ (mean value) and variance $V[m]$ [g$^{2}$] of the mass distribution.
    These are given by
    \begin{equation}
        E[m] = \frac{\rho_{\rm c}}{N_{\rm c}},
    \end{equation}
    and
    \begin{equation}
        V[m] = \frac{Z_{\rm c}}{N_{\rm c}} - \left(\frac{\rho_{\rm c}}{N_{\rm c}}\right)^{2}.
    \end{equation}
    In these examples, for simplicity we assume a particle density of $\rho_{\rm d}$ = 1 g cm$^{-3}$, the total number density is $N_{\rm c}$ = 1 cm$^{-3}$, the mass expectation value (mean) $E[m]$ = 10$^{-11}$ g  and particle size variance $V[m]$ = 10$^{-11}$ g$^{2}$.
    In Figure \ref{fig:dist} we show each of the derived distributions in mass and radius space.

    The mass grid, $m$ [g], is converted to a radius grid, $r$ [cm], assuming spherical particles from the relation
    \begin{equation}
        m = \frac{4}{3}\pi r^{3}\rho_{\rm d}.
    \end{equation}
    The mass particle size distribution, $f(m)$ [cm$^{-3}$ g$^{-1}$], is converted to the particle radius distribution, $f(r)$ [cm$^{-3}$ cm$^{-1}$], through the relation
    \begin{equation}
       f(r) =  \frac{3mf(m)}{r},
    \end{equation}
    where the factor of 3 is required to recover the correct integrated values of the distribution.
    Therefore, we can use the mass moments to fit a particle mass distribution, then convert it to a radius distribution to facilitate calculations.
    For the log-normal distribution, the median particle mass, $m_{\rm m}$ [g] and standard deviation, $\sigma_{\rm m}$ [g] are derived from the expectation value and variance from
    \begin{equation}
        \sigma_{\rm m} = \exp\left[\sqrt{\ln\left(1 + \frac{V[m]^{2}}{E[m]^{2}}\right)}\right],
    \end{equation}
    and
    \begin{equation}
        m_{\rm med} =  \frac{E[m]}{\exp\left[\ln(\sigma^{2})\right]}.
    \end{equation}
    The particle mass distribution is then
    \begin{equation}
        f(m) = \frac{N_{\rm c}}{m\sigma_{\rm m}\sqrt{2\pi}} \exp\left[-\frac{\left(\ln m - \ln m_{\rm med}\right)^{2}}{2\sigma_{\rm m}^{2}}\right].
    \end{equation}
    A useful relation between the standard deviation for the mass distribution and radius distribution for the log-normal distribution is
    \begin{equation}
        \sigma_{\rm r} = \frac{\sigma_{\rm m}}{3}.
    \end{equation}
    The mass weighted radius and median radius relationship for the log-normal distribution is given by
    \begin{equation}
      r_{\rm med} = r_{\rm c} \exp\left[-\frac{7}{2} \sigma_{\rm r}^2\right],
    \end{equation}
    This enables a simple way to convert the mean mass values from the moment scheme into radiative-transfer codes that generally require the radius distribution (e.g. gCMCRT).
    
    The inverse gamma distribution is given through first driving the $\alpha$ and $\beta$ parameters
    \begin{equation}
        \alpha = \frac{E[m]^{2}}{V[m]} + 2,
    \end{equation}
    \begin{equation}
        \beta = E[m](\alpha - 1).
    \end{equation}
    The mass distribution is then
    \begin{equation}
        f(m) = \frac{N_{\rm c}\beta^\alpha}{\Gamma(\alpha)} \left(\frac{1}{m}\right)^{\alpha+1} \exp(-\beta/m).
    \end{equation}

    The exponential distribution is a single parameter distribution given by the $\lambda$ parameter
    \begin{equation}
        \lambda = E[m],
    \end{equation}
    with the distribution given by
    \begin{equation}
        f(m) = \frac{N_{\rm c}}{\lambda}\exp(-m/\lambda).
    \end{equation}
    The exponential distribution is useful for consideration when using the 2-moment scheme as only the expectation value is required to estimate the $\lambda$ parameter.
    The relationship for the exponential distribution between the mass weighted radius and mean radius is given by
    \begin{equation}
        \langle r\rangle = \Gamma\left(\frac{4}{3}\right)r_{\rm c} \approx 0.893 r_{\rm c},
    \end{equation}
    where $\Gamma$ is the gamma function. 

    The Rayleigh distribution is another single parameter distribution given by the $\sigma$ value
    \begin{equation}
        \sigma = \frac{E[m]}{\sqrt{\pi/2}},
    \end{equation}
    where the distribution as
    \begin{equation}
        f(m) = \frac{N_{\rm c}}{\sigma^{2}}\exp\left(-\frac{m^{2}}{2\sigma^{2}}\right).
    \end{equation}
    As with the exponential distribution, as a single parameter distribution, the Rayleigh distribution is also useful to consider for 2-moment schemes.

    We note, that in our current scheme it is not possible (without approximation) to estimate other particle size distributions that require higher moment powers such as the potential exponential in \citet{Helling2008} and the Hansen distribution \citep{Hansen1971}.

    \subsection{Distribution integrated opacity}
    To calculate the extinction opacity, single scattering albedos and asymmetry factor from the particle mass distribution, the particle radius distribution must be first calculated as in the previous sections.
    Typically, the area weighted radius, known as the effective radius, $r_{\rm eff}$ [cm], can be used as a representative value to calculate the opacity.
    If the particle size distribution is well peaked, then $r_{\rm eff}$ $\approx$ $r_{\rm c}$ and Eq. \ref{eq:m_c} can be used in the opacity calculation.
    However, if the distribution is broad, then the effective radius can be usually calculated from considering the shape of the distribution function.
    For example, if the distribution is log-normal, the effective radius is related to the mass weighted radius from
    \begin{equation}
        r_{\rm eff} = r_{\rm c} \exp\left[-\sigma_{\rm r}^{2}\right].
    \end{equation}

    To calculate the distribution integrated quantities self-consistently, the following equations are used for the extinction opacity, $\bar{\kappa}_{\rm ext}$ [cm$^{2}$ g$^{-1}$],
    \begin{equation}
        \bar{\kappa}_{\rm ext} = \frac{1}{\rho_{\rm a}}\int_{0}^{\infty}\pi r^{2}Q_{\rm ext}(r)f(r)dr,
    \end{equation}
    the single scattering albedo, $\bar{\omega}$,
    \begin{equation}
        \bar{\omega} = \frac{\int_{0}^{\infty}\pi r^{2}Q_{\rm sca}(r)f(r)dr}{\int_{0}^{\infty}\pi r^{2}Q_{\rm ext}(r)f(r)dr},
    \end{equation}
    and asymmetry factor, $\bar{g}$, 
    \begin{equation}
        \bar{g} = \frac{\int_{0}^{\infty}\pi r^{2}Q_{\rm sca}(r)g(r)f(r)dr}{\int_{0}^{\infty}\pi r^{2}Q_{\rm sca}(r)f(r)dr}.
    \end{equation}
    The extinction and scattering efficiency factors $Q_{\rm ext}$$(r)$ and $Q_{\rm sca}$$(r)$ as well as the radius dependent asymmetry factor $g(r)$ can be calculated using Mie theory, approximations to Mie theory or other methods.

\end{appendix}

\label{LastPage}
\end{document}